\documentclass{jpp}
\usepackage{graphicx}
\usepackage{epstopdf, epsfig, enumitem,amsmath}
\usepackage{color}

\newcommand{\ba}{\textit{\textbf{a}}}
\newcommand{\bb}{\textit{\textbf{b}}}

\newcommand{\bj}{\textit{\textbf{j}}}
\newcommand{\bk}{\textit{\textbf{k}}}
\newcommand{\bp}{\textit{\textbf{p}}}
\newcommand{\bq}{\textit{\textbf{q}}}

\newcommand{\bu}{\textit{\textbf{u}}}
\newcommand{\bx}{\textit{\textbf{x}}}
\newcommand{\by}{\textit{\textbf{y}}}
\newcommand{\bz}{\textit{\textbf{z}}}
\newcommand{\bw}{{\boldsymbol{\omega}}}

\newcommand{\ii}{\mathrm{i}}

\definecolor{rodion}{RGB}{250,100,100}

\shorttitle{On uniqueness of transfer rates in MHD turbulence}
\shortauthor{F. Plunian, R. Stepanov and M.K. Verma}

\title{On uniqueness of transfer rates in magnetohydrodynamic turbulence}

\author{Franck Plunian\aff{1}
  \corresp{\email{Franck.Plunian@univ-grenoble-alpes.fr}},
 Rodion Stepanov\aff{2}
 \and 
 Mahendra Kumar Verma\aff{3}}

\affiliation{\aff{1}Universit\'e Grenoble Alpes, Universit\'e Savoie Mont Blanc, CNRS, IRD, IFSTTAR, ISTerre, 38000 Grenoble, France
\aff{2} Perm National Research Polytechnic University, Institute of Continuous Media Mechanics,  Korolyov 1,
Perm, 614013, Russia
\aff{3}Department of Physics, Indian Institute of Technology, Kanpur 208016, India}

\begin{document}

\maketitle

\begin{abstract}
In hydrodynamic and MHD (magnetohydrodynamic) turbulence, formal expressions for the transfer rates  rely on integrals over wavenumber triads $(\bk,\bp,\bq)$ satisfying $\bk+\bp+\bq=0$. As an example $S_E^{uu}(\bk|\bp,\bq)$ denotes the kinetic energy transfer rate to the mode $\bk$, from the two other modes in the triad, $\bp$ and $\bq$. However as noted by \cite{kraichnan1958}, in $S_E^{uu}(\bk|\bp,\bq)$, what fraction of the energy transferred to the mode $\bk$ originated from $\bp$ and which from $\bq$ is unknown . Such an expression is thus incongruent with the customary description of turbulence in terms of two-scale energy exchange. Notwithstanding this issue, \cite{DAR2001} further decomposed these transfers into separate contributions from $\bp$-to-$\bk$ and $\bq$-to-$\bk$, thus introducing the concept of mode-to-mode transfers that they applied to MHD turbulence. Doing so, they had to set aside additional transfers circulating within each triad, but failed to calculate them. 

In the present paper we explain how to derive the complete expressions of the mode-to-mode transfers, including the circulating transfers. We do it for kinetic energy and kinetic helicity in hydrodynamic turbulence, for kinetic energy, magnetic energy and magnetic helicity in MHD turbulence. We find that the degree of non-uniqueness of the energy transfers derived from the induction equation is a priori higher than the one derived from the Navier-Stokes equations. However separating the contribution of magnetic advection from magnetic stretching, the energy mode-to-mode transfer rates involving the magnetic field become uniquely defined, in striking contrast to the hydrodynamic case. The magnetic helicity mode-to-mode transfer rate is also found to be uniquely defined, contrary to kinetic helicity in hydrodynamics. We find that shell-to-shell transfer rates have the same properties as mode-to-mode transfer rates. Finally calculating the fluxes, we show how the circulating transfers cancel in accordance with conservation laws. 
\end{abstract}


\section{Introduction}
In hydrodynamic and MHD (magnetohydrodynamic) homogeneous turbulence, the energy exchange between scales is often introduced as two-scale transfers. These transfers can be either local or non-local, depending if in Fourier space the two corresponding wavenumbers belong to neighbouring or distant shells. 
In hydrodynamic turbulence, the classic picture of \cite{Richardson1922} corresponds to local transfers, the energy cascading towards smaller and smaller scales.
In MHD turbulence, the motion of an electroconducting fluid can generate a magnetic field by the so-called dynamo effect, with a magnetic scale much larger than the motion scale \citep{Stieglitz2001,Muller2004}. In this case the energy transfer is non-local, from the flow small scale to the magnetic large scale. 

This two-scale picture of energy transfers is in fact misleading as these transfers result from quadratic nonlinearities involving, in Fourier space, not two but three wavenumbers $(\bk,\bp,\bq)$, satisfying $\bk+\bp+\bq=0$. As an example let us consider the kinetic energy equation in the hydrodynamic case. At wavenumber $\bk$ the kinetic energy satisfies the following equation
\begin{equation}
( \partial_t  + 2\nu k^2) E^u_\bk  = \frac{1}{2}\sum \limits_{\underset{\bf k+p+q=0}{\bp,\bq}}
S_E^{uu}\left(\bk|\bp,\bq\right)
\end{equation}
where $\nu $ is the viscosity and $k=|\bk|$. On the right hand side $S_E^{uu}(\bk|\bp,\bq)$ denotes the kinetic energy transfer rate  received by the mode $\bk$, coming from the two other modes of the triad $\bp$ and $\bq$. 
It is defined by
\begin{equation}
\label{eq:Suu}
S_E^{uu}\left(\bk|\bp,\bq\right)=\Imag\{(\bp\cdot\bu_\bq)(\bu_\bp\cdot\bu_\bk)+(\bq\cdot\bu_\bp)(\bu_\bq\cdot\bu_\bk)\},
\end{equation}
where $\bu_\bk, \bu_\bp$ and $\bu_\bq$ are the Fourier coefficients of $\bu$ at wavenumbers $\bk, \bp$ and $\bq$.
Obviously $S_E^{uu}(\bk|\bp,\bq)$ is symmetric with respect to $\bp$ and $\bq$, then there is a priori no possibility to distinguish what fraction of the energy transferred to the mode $\bk$ originated from $\bp$ and which from $\bq$ \citep{kraichnan1958,kraichnan1959}.

To circumvent this problem one may integrate $S_E^{uu}(\bk|\bp,\bq)$ on two Fourier shells $P$ and $K$, $\bp$ belonging to the giver shell $P$ and $\bk$ belonging to the receiver shell $K$, with $\bq=-(\bk+\bp)$. Such a concept of shell-to-shell transfer rate has long been investigated in hydrodynamic turbulence \citep{Batchelor1953} and more recently in MHD \citep{Alexakis2005,Mininni2005,Mininni2011}. However it does not tell us to which shell $\bq$ belongs to.  In particular, the fact that $\bq$ may belong to neither shell $P$ or $K$, makes unreliable the determination of  the degree of non-locality of the transfers  \citep{Domaradzki1990,Domaradzki1992}. 
Another approach, based on spatial coarse graining has shown dominant local transfers in both hydrodynamics and MHD turbulence cases \citep{Aluie2009,Aluie2010}.

Notwithstanding the symmetry issue raised by \cite{kraichnan1958}, \cite{DAR2001} further decomposed $S_E^{uu}\left(\bk|\bp,\bq\right)$ into separate contributions from $\bp$-to-$\bk$ and $\bq$-to-$\bk$. They denoted $S_E^{uu}\left(\bk|\bp|\bq\right)$ the mode-to-mode transfer rate of kinetic energy coming from $\bu_\bp$ towards $\bu_\bk$, with $\bu_\bq$ acting as a mediator. Such mode-to-mode transfer rate must satisfy the following two conditions
\begin{eqnarray}
\label{eq:Suumtm1}
S_E^{uu}\left(\bk|\bp,\bq\right)&=&S_E^{uu}\left(\bk|\bp|\bq\right)+S_E^{uu}\left(\bk|\bq|\bp\right)\\
S_E^{uu}\left(\bk|\bp|\bq\right)&=&-S_E^{uu}\left(\bp|\bk|\bq\right),
\label{eq:Suumtm2}
\end{eqnarray} 
the second equation meaning that both mode-to-mode transfers $\bu_\bp$-to-$\bu_\bk$ and $\bu_\bk$-to-$\bu_\bp$, with the same mediator $\bu_\bq$, are opposite.
On the right hand side of (\ref{eq:Suu}) they ascribed the first term to $S_E^{uu}\left(\bk|\bp|\bq\right)$ and the second one to $S_E^{uu}\left(\bk|\bq|\bp\right)$
\begin{subeqnarray}
\label{eq:Tuu1}
S_E^{uu}\left(\bk|\bp|\bq\right)&=&\Imag\{(\bp\cdot\bu_\bq)(\bu_\bp\cdot\bu_\bk)\}\\
S_E^{uu}\left(\bk|\bq|\bp\right)&=&\Imag\{(\bq\cdot\bu_\bp)(\bu_\bq\cdot\bu_\bk)\}.
\end{subeqnarray}
This definition has been applied to hydrodynamics \citep{Verma2005} and extended to MHD turbulence  \citep{DAR2001,Verma2004,Carati2006,Teaca2009a,Teaca2009b}, convection \citep{Kumar2014} and elastic wave turbulence \citep{Yokoyama2017}. 

Now, anticipating the next section it can be shown that the expression given in (\ref{eq:Suu}) is exactly equivalent to the following one  
\begin{equation}
\label{eq:Suu2}
S_E^{uu}\left(\bk|\bp,\bq\right)=\Real\{(\bu_\bk,\bu_\bp,\bw_\bq)+(\bu_\bk,\bu_\bq,\bw_\bp)\},
\end{equation}
where $\Real\{(\bx,\by,\bz)\}$ denotes the real part of the mixed product of the  complex vectors $\bx, \by$ and $\bz$, and with $\bw_\bp$ and  $\bw_\bq$ the Fourier coefficients of vorticity at wavenumbers $\bp$ and $\bq$.
Accordingly, another definition of mode-to-mode transfers satisfying (\ref{eq:Suumtm1}) and (\ref{eq:Suumtm2}) can be derived
\begin{subeqnarray}
\label{eq:Tuu2}
S_E^{uu}\left(\bk|\bp|\bq\right)&=&\Real\{(\bu_\bk,\bu_\bp,\bw_\bq)\}\\
S_E^{uu}\left(\bk|\bq|\bp\right)&=&\Real\{(\bu_\bk,\bu_\bq,\bw_\bp)\}.
\end{subeqnarray}
Both definitions (\ref{eq:Tuu1}) and (\ref{eq:Tuu2}) are not equivalent and may lead to different results, e.g. in terms of locality of the transfers. 
Yet there is no reason to choose one over the other.

As will be shown below, this paradox is in fact related to the non-uniqueness of mode-to-mode transfers.
We will see in section \ref{sec:enmtm} that the full expression of the mode-to-mode transfer  is actually given by
\begin{equation}
T_{E}(\bu_\bp\overset{\bu_\bq}{\longrightarrow}\bu_\bk)=\Imag\{(\bp\cdot\bu_\bq)(\bu_\bp\cdot\bu_\bk)\}+\Delta_E^u(\bk|\bp|\bq),
\label{eq:SEuu}
\end{equation}
with
\begin{equation}
\Delta_E^u(\bk|\bp|\bq)=\alpha_E^u\Real\{(\bu_{\bk}, \bu_{\bp}, \bw_{\bq})
+ (\bu_{\bk}, \bw_{\bp}, \bu_{\bq})
+(\bw_{\bk}, \bu_{\bp}, \bu_{\bq})\}, \label{eq:Delta}
\end{equation}
$\alpha_E^u$ being an arbitrary real coefficient. 
In (\ref{eq:SEuu}) we adopt a new notation for the mode-to-mode transfer in order to differentiate it from the definition (\ref{eq:Tuu1}) of \cite{DAR2001}, the latter being a particular case  of (\ref{eq:SEuu}) corresponding to $\alpha_E^u=0$.
The alternative definition given in (\ref{eq:Tuu2}) corresponds to $\alpha_E^u=1/2$.

The first objective of the present paper is to explain how to derive expressions (\ref{eq:SEuu}) and (\ref{eq:Delta}), but also to generalize them for kinetic helicity in hydrodynamics, for kinetic energy, magnetic energy and magnetic helicity in MHD. For brevity, we will directly consider the MHD case, the hydrodynamic case being obtained by cancelling the magnetic field.
Starting from the Navier-Stokes and induction equations projected onto Fourier space (section \ref{sec:generalMHD}), we will derive a basis of functions for each type of transfer in order to obtain the complete expression for each mode-to-mode transfer rate (section \ref{sec:modetomode}). 
Finally we will address the issue of shell-to-shell transfer rates and fluxes in the light of the new complete expressions of mode-to-mode transfer rates (section \ref{sec:conclusion}).

\section{Energy and helicity transfer rates, from $\bp$ and $\bq$, to $\bk$}
\label{sec:generalMHD}
\subsection{MHD equations in Fourier space}
\label{sec:MHD equations}
\begin{figure*}
\centering
\includegraphics[width=0.5\columnwidth]{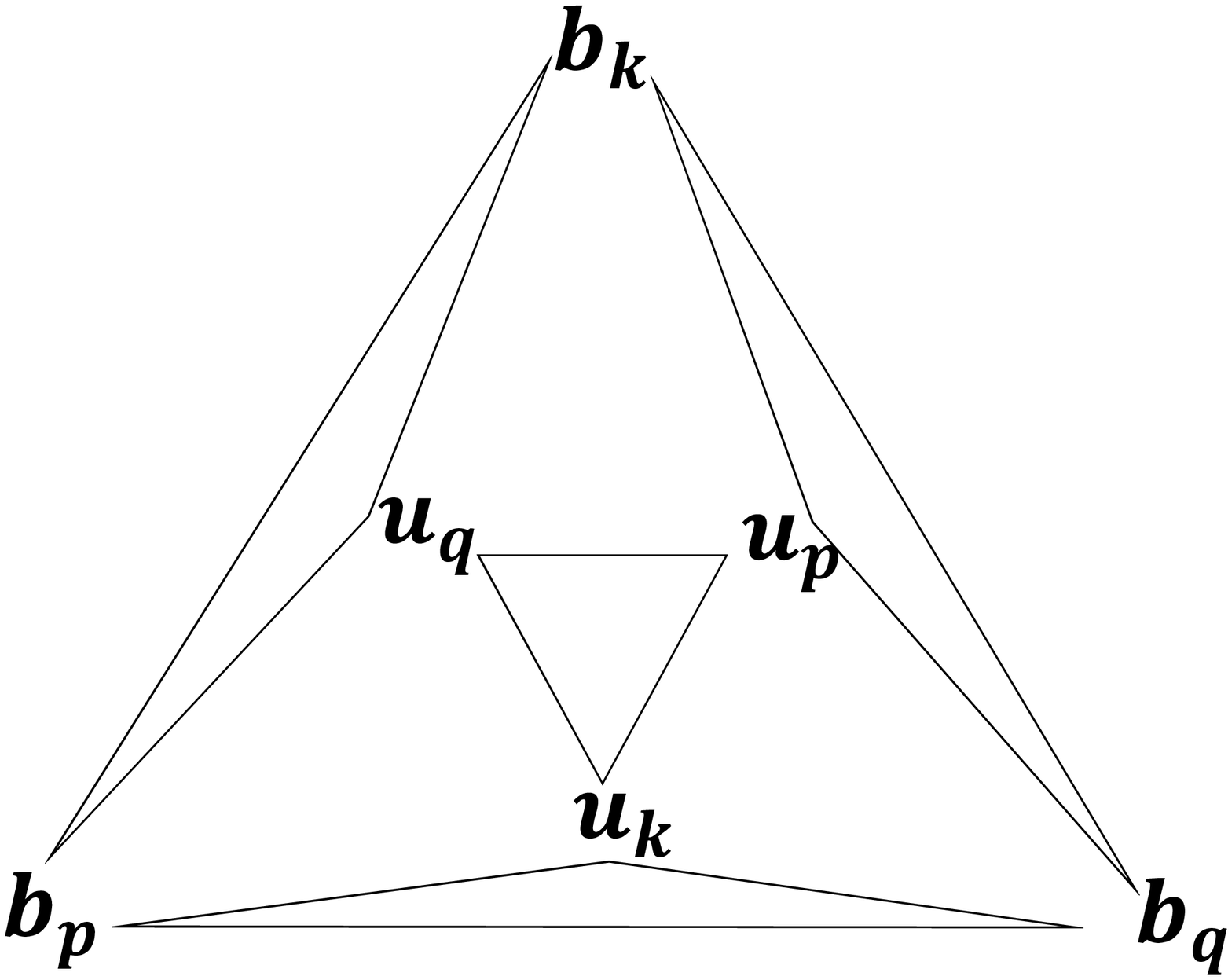}
\caption{\label{fig: MHD triad} Within one triad $(\bk,\bp,\bq)$ the triangles show all possible nonlinear interactions in MHD turbulence, between either the three velocity modes, or two magnetic modes and one velocity mode. In hydrodynamic turbulence ($\bb=0$) the nonlinear interactions occur only between the three velocity modes.}
\end{figure*}

The general equations for MHD for an incompressible and electrically conducting fluid are given by 
\footnote{Another form of the Navier-Stokes equation can be obtained from (\ref{eq:u1}) using 
 the identity $-(\bu\cdot\nabla)\bu= \bu \times \bw - \nabla (\bu^2/2)$.}:
\begin{eqnarray}
\partial_t \bu -  \nu\nabla^2 {\bu} & = & \bu\times\bw+ 
\bj \times \bb -\nabla \left(p+\bu^2/2\right), 
\label{eq:u1}\\
\partial_t \bb  - 
\eta\nabla^2 {\bb} & = & \nabla \times\left(\bu \times \bb\right), \label{eq:b1}\\
\nabla \cdot \bu  &=& \nabla \cdot \bb =  0, \label{eq:div}
\end{eqnarray}
where $\bu$ denotes the fluid velocity, $\bb$ the magnetic induction normalized by the square root of the fluid density times the fluid magnetic permeability, $p$ the pressure field normalized by the fluid density, $\nu$ and $\eta$ the kinetic viscosity and magnetic diffusivity. The vorticity $\bw$ and the normalized current density $\bj$ are respectively defined by 
\begin{equation}
\bw=\nabla \times \bu,\;\;\;\;\; \bj=\nabla \times \bb. 
\end{equation}
Assuming periodic boundary conditions, we can decompose the velocity, pressure and magnetic fields in Fourier series,
leading to the following equations for the Fourier coefficients 
\begin{eqnarray}
(\partial_t + \nu k^2)\bu_\bk  &=&   \sum \limits_{\underset{\bf k+p+q=0}{\bp,\bq}} \left(\bu_\bp^*\times\bw_\bq^* + \bj_\bp^*\times\bb_\bq^*\right)- \ii \left(p_\bk+\frac{(\bu^2)_\bk}{2}\right)\bk, \label{eq:u2}\\
(\partial_t + \eta k^2)\bb_\bk  &=& \sum \limits_{\underset{\bf k+p+q=0}{\bp,\bq}} \ii\bk\times \left(\bu_\bp^*\times\bb_\bq^*\right), \label{eq:b2}\\
{\bk \cdot \bu_\bk}&=&{\bk \cdot \bb_\bk}  =  0,
\end{eqnarray}
with $k=|\bk|$, 
where the asterisk denotes the complex conjugation, and where the vorticity and current density satisfy
\begin{equation}
\bw_\bk=\ii\bk\times\bu_\bk,\;\;\;\;\bj_\bk=\ii\bk\times\bb_\bk.
\end{equation}
In (\ref{eq:u2}) and (\ref{eq:b2}) the complex conjugations come from the fact that $\bk=-\bp -\bq$ and that any Fourier coefficient must satisfy  
$\bx_{-\bk}=\bx_{\bk}^*$ in order to guarantee that $\bu$, $\bb$, $\bw$ and $\bj$ are real vectors.
The sum $\sum \limits_{\underset{\bf k+p+q=0}{\bp,\bq}}$ means the double sum on all $\bp$ and $\bq$ satisfying $\bk+\bp+\bq=0$, which is equivalent to the sum on all $\bp$ with $\bq=-(\bp+\bk)$ and equivalent to the sum on all $\bq$ with $\bp=-(\bq+\bk)$. Within a triad $(\bk,\bp,\bq)$ the possible nonlinear interactions  are summarized in figure \ref{fig: MHD triad}.
\subsection{Kinetic and magnetic energy transfer rates}
\label{sec:enequation}
In Fourier space, the kinetic and magnetic energies are defined by
\begin{equation}
E^u_\bk =  \frac{1}{2} \bu_\bk\cdot\bu_\bk^*,\;\;\;
E^b_\bk =  \frac{1}{2} \bb_\bk\cdot\bb_\bk^*.\;\;\;
\label{Energy_Fourier}
\end{equation}
Taking advantage of the fact that (\ref{eq:u2}) and (\ref{eq:b2}) are invariant under the exchange of $\bp$ and $\bq$, the kinetic and magnetic energies
satisfy the following equations \footnote{In (\ref{Energy_Fourier_timeu2}-\ref{Energy_Fourier_timeb2}) the identity $\sum \limits_{\underset{\bf k+p+q=0}{\bp,\bq}}\bx_\bp \times \by_\bq= \frac{1}{2}\sum \limits_{\underset{\bf k+p+q=0}{\bp,\bq}}\left(\bx_\bp \times \by_\bq+\bx_\bq \times \by_\bp\right)$ has been applied.}:
\begin{eqnarray}
( \partial_t  + 2\nu k^2) E^u_\bk  &=& \frac{1}{2}\sum \limits_{\underset{\bf k+p+q=0}{\bp,\bq}}
S_E^{uu}\left(\bk|\bp,\bq\right)+S_E^{ub}\left(\bk|\bp,\bq\right),\label{Energy_Fourier_timeu2}\\
( \partial_t  + 2\eta k^2) E^b_\bk  &=& \frac{1}{2}\sum \limits_{\underset{\bf k+p+q=0}{\bp,\bq}}
S_E^{bu}\left(\bk|\bp,\bq\right)+S_E^{bb}\left(\bk|\bp,\bq\right),\label{Energy_Fourier_timeb2}
\end{eqnarray}
with
\begin{eqnarray}
S_E^{uu}\left(\bk|\bp,\bq\right)=&\Real\{(\bu_\bk,\bu_\bp,\bw_\bq)&-(\bu_\bk,\bw_\bp,\bu_\bq)\},\label{SE1}\\
S_E^{ub}\left(\bk|\bp,\bq\right)=&\Real\{(\bu_\bk,\bj_\bp,\bb_\bq)&-(\bu_\bk,\bb_\bp,\bj_\bq)\},\label{SE2}\\
S_E^{bu}\left(\bk|\bp,\bq\right)+S_E^{bb}\left(\bk|\bp,\bq\right)=&\Real\{(\bj_\bk,\bu_\bp,\bb_\bq)&-(\bj_\bk,\bb_\bp,\bu_\bq)\}.\label{SE3}
\end{eqnarray}
Following  \cite{DAR2001}, $S_E^{xy}\left(\bk|\bp,\bq\right)$ is understood as the transfer rate of $y$-energy at wavenumbers $\bp$ and $\bq$, to $x$-energy at wavenumber $\bk$. By definition $S_E^{xy}\left(\bk|\bp,\bq\right)$ must be symmetric with respect to $\bp$ and $\bq$,
\begin{equation}
S_E^{xy}\left(\bk|\bp,\bq\right)=S_E^{xy}\left(\bk|\bq,\bp\right).
\end{equation}

At this stage we make several remarks. 

First, we note that the definition (\ref{SE1}) of
$S_E^{uu}\left(\bk|\bp,\bq\right)$, which has already been given in (\ref{eq:Suu2}), also coincides with the definition given in (\ref{eq:Suu}). This can be shown by applying the following identity twice
\begin{equation} \label{identity1}
-\ii(\bp\cdot\bu_{\bq})( \bu_{\bp}\cdot\bu_{\bk})=\frac{1}{2}[\left(\bu_{\bk}, \bu_{\bp}, \bw_{\bq}\right) -\left(\bu_{\bk}, \bw_{\bp}, \bu_{\bq}\right)  - \left(\bw_{\bk}, \bu_{\bp}, \bu_{\bq}\right)],
\end{equation}
once as it is, once by exchanging $\bp$ and $\bq$.
The demonstration of identity (\ref{identity1}) is given in Appendix \ref{Appsec:identity}.

Second, the definition (\ref{SE2}) of the transfer rate of magnetic energy at wavenumbers $\bp$ and $\bq$ to kinetic energy at wavenumber $\bk$, $S_E^{ub}\left(\bk|\bp,\bq\right)$, coincides with the definition given in \cite{DAR2001}.

Third, contrary to $S_E^{uu}\left(\bk|\bp,\bq\right)$ and $S_E^{ub}\left(\bk|\bp,\bq\right)$ which are uniquely defined, $S_E^{bu}\left(\bk|\bp,\bq\right)$ and $S_E^{bb}\left(\bk|\bp,\bq\right)$ cannot be inferred from the only one equation (\ref{SE3}). Nevertheless, in the induction equation (\ref{eq:b1}) there is some relevance to separating the nonlinear term into an advection term and a stretching term, which can give insights into the dynamics of the system. In that case $S_E^{bu}\left(\bk|\bp,\bq\right)$ and $S_E^{bb}\left(\bk|\bp,\bq\right)$ are uniquely defined. We will come back on this important issue at the end of section \ref{sec:enmtm}. 

Finally, (\ref{SE1}), (\ref{SE2}) and (\ref{SE3}) imply the following relations
\begin{eqnarray}
S_E^{uu}\left(\bk|\bp,\bq\right)&+&
S_E^{uu}\left(\bp|\bq,\bk\right)+
S_E^{uu}\left(\bq|\bk,\bp\right)=0,\label{SumSuuu}\\
\left(S_E^{ub}+S_E^{bu}+S_E^{bb}\right)\left(\bk|\bp,\bq\right)&+&
\left(S_E^{ub}+S_E^{bu}+S_E^{bb}\right)\left(\bp|\bq,\bk\right)\nonumber\\
&+&\left(S_E^{ub}+S_E^{bu}+S_E^{bb}\right)\left(\bq|\bk,\bp\right)=0.\label{SumSubb}
\end{eqnarray}
In hydrodynamics (\ref{SumSuuu}) leads to the {\it detailed conservation of kinetic energy} \citep{kraichnan1959}, meaning that in each triad $(\bk,\bp,\bq)$ and in the inviscid limit $\nu=0$, the kinetic energy is conserved, i.e.
\begin{equation}
\label{eq:consEu}
\partial_t\left(E_\bk^u+E_\bp^u+E_\bq^u\right)=0.
\end{equation}
In MHD (\ref{SumSuuu}) and (\ref{SumSubb}) imply that in each triad $(\bk,\bp,\bq)$ and in the inviscid and diffusionless limits $\nu=\eta=0$, the sum of kinetic and magnetic energies is conserved, i.e.
\begin{equation}
\label{eq:consEtot}
\partial_t\left(E_\bk^u+E_\bk^b+E_\bp^u+E_\bp^b+E_\bq^u+E_\bq^b\right)=0,
\end{equation}
which states the {\it detailed conservation of total energy}.

\subsection{Kinetic and magnetic helicity transfer rates}
\label{sec:helequation}
In Fourier space, the kinetic and magnetic helicities are defined by
\begin{equation}
H^u_\bk =  \frac{1}{2} \bu_\bk\cdot\bw_\bk^*,\;\;\;\;\;\;
H^b_\bk =  \frac{1}{2} \bb_\bk\cdot\ba_\bk^*,
\label{Helicity_Fourier}
\end{equation}
where $\ba_\bk$ is the Fourier coefficient of the potential vector, satisfying
\begin{equation}
\bb_\bk=\ii\bk\times \ba_\bk.
\end{equation}
Taking advantage again of the fact that (\ref{eq:u2}) and (\ref{eq:b2}) are invariant under the exchange of $\bp$ and $\bq$, the kinetic and magnetic helicities
satisfy the following equations
\begin{eqnarray}
( \partial_t  + 2\nu k^2) H^u_\bk  &=& \frac{1}{2}\sum \limits_{\underset{\bf k+p+q=0}{\bp,\bq}}
S_H^u\left(\bk|\bp,\bq\right)+F_H^u\left(\bk|\bp,\bq\right),\label{Helicity_Fourier_timeu2}\\
( \partial_t  + 2\eta k^2) H^b_\bk  &=& \frac{1}{2}\sum \limits_{\underset{\bf k+p+q=0}{\bp,\bq}}
S_H^b\left(\bk|\bp,\bq\right),\label{Helicity_Fourier_timeb2}
\end{eqnarray}
with
\begin{eqnarray}
S_H^u\left(\bk|\bp,\bq\right)&=&\Real\{(\bw_\bk,\bu_\bp,\bw_\bq)-(\bw_\bk,\bw_\bp,\bu_\bq)\},\label{SH1}\\
F_H^u\left(\bk|\bp,\bq\right)&=&\Real\{(\bw_\bk,\bj_\bp,\bb_\bq)-(\bw_\bk,\bb_\bp,\bj_\bq)\},\label{SH2}\\
S_H^b\left(\bk|\bp,\bq\right)&=&\Real\{(\bb_\bk,\bu_\bp,\bb_\bq)-(\bb_\bk,\bb_\bp,\bu_\bq)\}.\label{SH3}
\end{eqnarray}
Here, $S_H^u\left(\bk|\bp,\bq\right)$ denotes the kinetic helicity transfer rate from wavenumbers $\bp$ and $\bq$ to $\bk$, and satisfies the following identity:
\begin{equation}
S_H^u\left(\bk|\bp,\bq\right) +
S_H^u\left(\bp|\bq,\bk\right)+
S_H^u\left(\bq|\bk,\bp\right)=0.\label{SumSuuuhel}
\end{equation}
In hydrodynamics (\ref{SumSuuuhel}) leads to the {\it detailed conservation of kinetic helicity}, meaning that in each triad $(\bk,\bp,\bq)$ and in the inviscid limit $\nu=0$, the kinetic helicity is conserved i.e.
\begin{equation}
\partial_t\left(H_\bk^u+H_\bp^u+H_\bq^u\right)=0.
\label{conservhel}
\end{equation}
In MHD, even though (\ref{SumSuuuhel}) still holds, the kinetic helicity is not conserved anymore because the magnetic field produces an additional source of kinetic helicity $F_H^u\left(\bk|\bp,\bq\right)$ that does not satisfy any identity as above. 

In (\ref{SH3}) $S_H^b\left(\bk|\bp,\bq\right)$ denotes the magnetic helicity transfer rate from wavenumbers $\bp$ and $\bq$ to $\bk$, and satisfies the following identity
\begin{equation}
S_H^b\left(\bk|\bp,\bq\right) +
S_H^b\left(\bp|\bq,\bk\right)+
S_H^b\left(\bq|\bk,\bp\right)=0.\label{SumShelb}
\end{equation}
It leads to the {\it detailed conservation of magnetic helicity}, meaning that in each triad $(\bk,\bp,\bq)$ and in the diffusionless limit $\eta=0$, the magnetic helicity is conserved i.e.
\begin{equation}
\partial_t\left(H_\bk^b+H_\bp^b+H_\bq^b\right)=0.
\label{conservhelb}
\end{equation}

\section{Mode-to-mode transfer rates}
\label{sec:modetomode}
\subsection{Generic definition of mode-to-mode transfer rates}
\label{sec:defmodetomode}
Consider three vectors $\bx_\bk,\by_\bp,\bz_\bq$ such that $\bx,\by$ and $\bz$ denote either three velocity fields, or one velocity field and two magnetic fields, as illustrated in figure \ref{fig: MHD triad}. Again we assume that the wavenumber triad $(\bk,\bp,\bq)$ satisfies $\bk+\bp+\bq=0$.

Consider a generic transfer rate $S_{E,H}^{xy}(\bk|\bp,\bq)$ which can be a transfer rate of either energy or helicity with the appropriate subscript $E$ or $H$. As $S_{E,H}^{xy}(\bk|\bp,\bq)$ is symmetric with respect to $\bp$ and $\bq$, we can decompose it in two parts 
\begin{eqnarray}
S_{E,H}^{xy}(\bk|\bp,\bq) &=& T_{E,H}(\by_\bp\overset{\bz_\bq}{\longrightarrow}\bx_\bk)+T_{E,H}(\by_\bq\overset{\bz_\bp}{\longrightarrow}\bx_\bk), \label{eq:SversusT}
\end{eqnarray}
where $T_{E,H}(\by_\bp\overset{\bz_\bq}{\longrightarrow}\bx_\bk)$ is interpreted as the mode-to-mode transfer rate of energy, or helicity, from the giver mode $\by_\bp$ to the receiver mode $\bx_\bk$, with the mode $\bz_\bq$ acting as a mediator. 
Moreover, using the same mediator mode $\bz_\bq$, both mode-to-mode transfer rates $\by_\bp$-to-$\bx_\bk$ and $\bx_\bk$-to-$\by_\bp$ must be opposite, 

\begin{eqnarray}
\label{eq:Topposite}
T_{E,H}(\by_\bp\overset{\bz_\bq}{\longrightarrow}\bx_\bk)&=&
-T_{E,H}(\bx_\bk\overset{\bz_\bq}{\longrightarrow}\by_\bp).
\end{eqnarray}
Again, we adopt another notation than the one given in \cite{DAR2001} because, as will be shown below, the latter is only a particular case of the one we are deriving here.

The two conditions (\ref{eq:SversusT}) and (\ref{eq:Topposite}) imply the following relation
\begin{equation}
(S_{E,H}^{xy}+S_{E,H}^{yx})(\bk|\bp,\bq)+
(S_{E,H}^{xy}+S_{E,H}^{yx})(\bp|\bq,\bk)+
(S_{E,H}^{xy}+S_{E,H}^{yx})(\bq|\bk,\bp)=0.
\label{eq:sumSxy}
\end{equation}
Taking $\bx=\by=\bu$, equation (\ref{eq:sumSxy}) corresponds to (\ref{SumSuuu}) for the kinetic energy transfers, and to (\ref{SumSuuuhel}) for the kinetic helicity transfers. 
For the MHD energy transfers and for respectively $\bx=\by=\bb$ and $\bx=\bu$, $\by=\bb$, equation (\ref{eq:sumSxy}) corresponds to
\begin{eqnarray}
S_E^{bb}(\bk|\bp,\bq)+
S_E^{bb}(\bp|\bq,\bk)+
S_E^{bb}(\bq|\bk,\bp)=0,
\label{eq:sumSEbb}\\
\left(S_E^{ub}+S_E^{bu}\right)\left(\bk|\bp,\bq\right)+
\left(S_E^{ub}+S_E^{bu}\right)\left(\bp|\bq,\bk\right)
+\left(S_E^{ub}+S_E^{bu}\right)\left(\bq|\bk,\bp\right)=0,\label{SumSubb2}
\end{eqnarray}
the sum of both leading to (\ref{SumSubb}).
For the MHD helicity transfers $\bx=\by=\bb$, (\ref{eq:sumSxy}) corresponds to (\ref{SumShelb}). 

\subsection{Kinetic and magnetic energy mode-to-mode transfer rates}
\label{sec:enmtm}
After (\ref{SE1}-\ref{SE3}), we can define a basis of functions in real space, $\Real\{(\bx_\bk,\by_\bp,\ii\bq\times\bz_\bq)\}$, $\Real\{(\bx_\bk,\ii\bp\times\by_\bp,\bz_\bq)\}$
and $\Real\{(\ii\bk\times\bx_\bk,\by_\bp,\bz_\bq)\}$,  such that $T_E(\by_\bp\overset{\bz_\bq}{\longrightarrow}\bx_\bk)$ is expressed as a linear combination of them. 
Then, the four energy mode-to-mode transfer rates take the following form
\begin{eqnarray}
T_E(\bu_\bp\overset{\bu_\bq}{\longrightarrow}\bu_\bk) &=& A_0\Real\{(\bu_{\bk}, \bu_{\bp}, \bw_{\bq})\}
+ B_0\Real\{(\bu_{\bk}, \bw_{\bp}, \bu_{\bq})\}
+C_0\Real\{(\bw_{\bk}, \bu_{\bp}, \bu_{\bq})\},\;\;\;\;\;\;\label{basisuu}\\T_E(\bb_\bp\overset{\bb_\bq}{\longrightarrow}\bu_\bk) &=& A_1\Real\{(\bu_{\bk}, \bb_{\bp}, \bj_{\bq})\}
+ B_1\Real\{(\bu_{\bk}, \bj_{\bp}, \bb_{\bq})\}
+C_1\Real\{(\bw_{\bk}, \bb_{\bp}, \bb_{\bq})\},\label{basisub}\\
T_E(\bb_\bp\overset{\bu_\bq}{\longrightarrow}\bb_\bk) &=& A_2\Real\{(\bb_{\bk}, \bb_{\bp}, \bw_{\bq})\}
+ B_2\Real\{(\bb_{\bk}, \bj_{\bp}, \bu_{\bq})\}
+C_2\Real\{(\bj_{\bk}, \bb_{\bp}, \bu_{\bq})\},\label{basisbb}\\
T_E(\bu_\bp\overset{\bb_\bq}{\longrightarrow}\bb_\bk) &=& A_3\Real\{(\bb_{\bk}, \bu_{\bp}, \bj_{\bq})\}
+ B_3\Real\{(\bb_{\bk}, \bw_{\bp}, \bb_{\bq})\}
+C_3\Real\{(\bj_{\bk}, \bu_{\bp}, \bb_{\bq})\},\label{basisbu}
\end{eqnarray}
where the coefficients $A_l,B_m, C_n$, $(l,m,n)\in [0,3]^3$ are scalar quantities.

\begin{figure*}
\centering
\includegraphics[width=0.5\columnwidth]{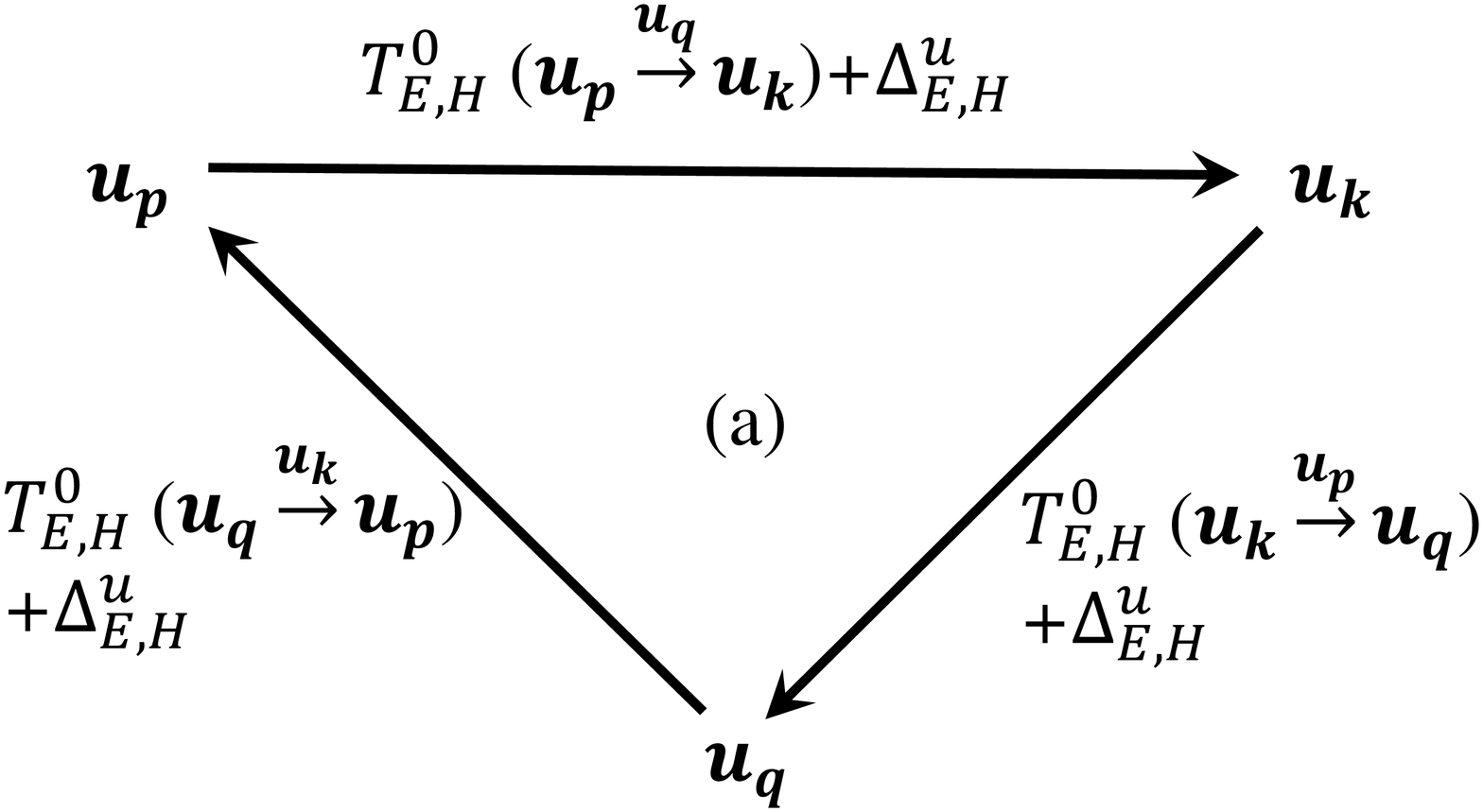}\hfill
\includegraphics[width=0.5\columnwidth]{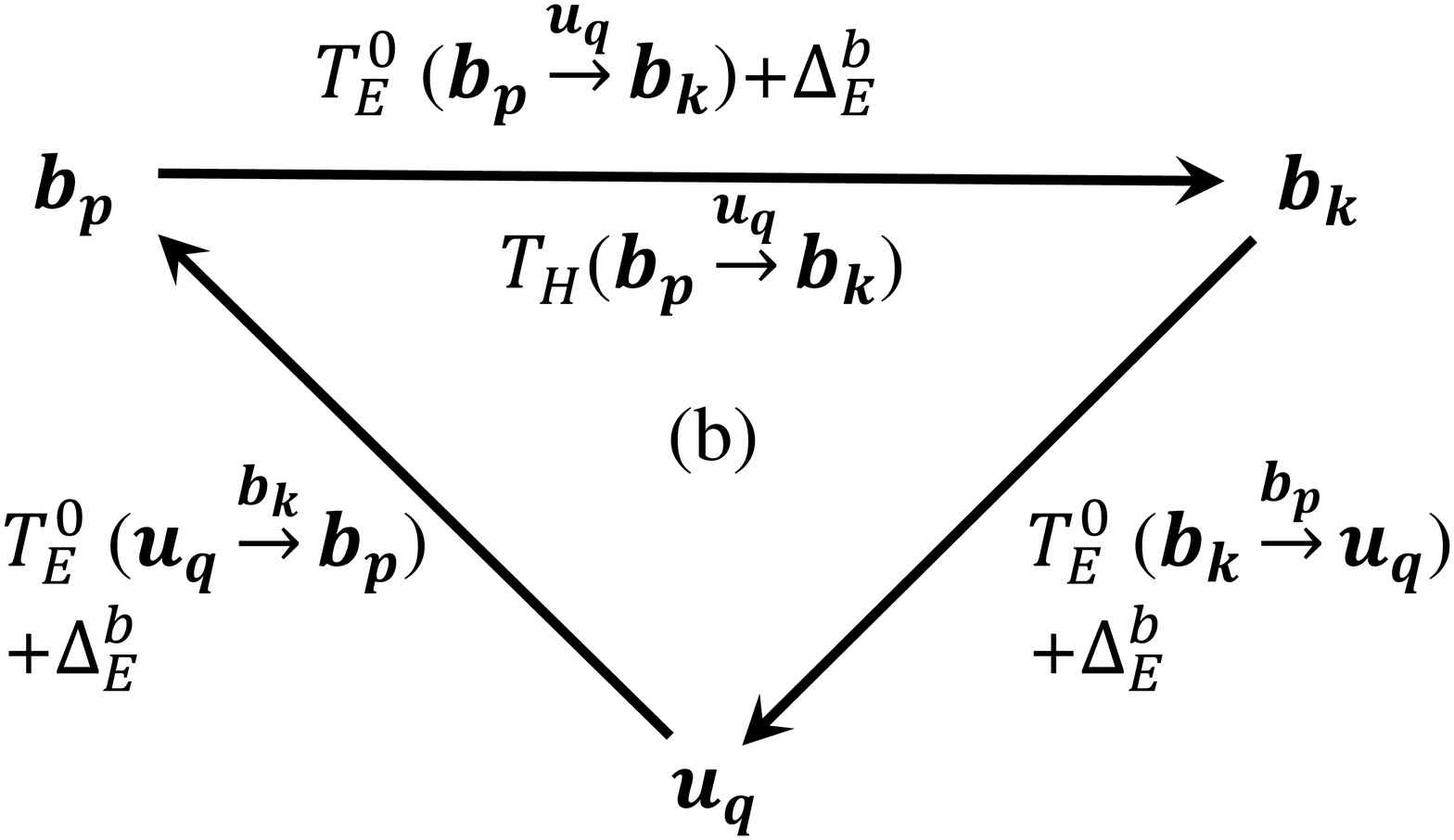}
\caption{\label{fig: Energy transfers} The mode-to-mode energy (subscript $E$) and helicity (subscript $H$) transfer rates occur within triplets of modes $(\bu_\bk,\bu_\bp,\bu_\bq)$ or $(\bb_\bk,\bb_\bp,\bu_\bq)$, illustrated respectively on left side (a) and right side (b).
Each energy mode-to-mode transfer rate is decomposed in two parts,   $T^0_E$ corresponding to the first term on the right hand side of (\ref{Tuu}-\ref{Tbu}), the circulating terms $\Delta_E^u$ and $\Delta_E^b$ being defined in (\ref{eq:DeltaEu}) and (\ref{eq:DeltaEb}).
In hydrodynamics each kinetic helicity mode-to-mode transfer rate is decomposed in two parts,  $T^0_H$ corresponding to the first term on the right hand side of (\ref{TuuH}), the circulating terms $\Delta_H^u$ being defined in (\ref{eq:DeltaHu}). In MHD the magnetic helicity mode-to-mode transfer rate is defined in (\ref{TbbH}).}
\end{figure*}

Applying the two conditions (\ref{eq:SversusT}) and (\ref{eq:Topposite}), together with (\ref{SE1}-\ref{SE3}), and making use of the following identity (see Appendix \ref{Appsec:identity})
\begin{equation} \label{identity2}
 \left(\bx_{\bk}, \ii\bp\times\by_{\bp}, \bz_{\bq}\right)  + \left(\ii\bk\times\bx_{\bk}, \by_{\bp}, \bz_{\bq}\right)=\left(\bx_{\bk}, \by_{\bp}, \ii \bq \times\bz_{\bq}\right)+ 2\ii(\bp\cdot\bz_{\bq})( \by_{\bp}\cdot\bx_{\bk})
\end{equation}
which is satisfied provided $\bk+\bp+\bq=0$ and $\bk\cdot\bx_\bk=\bp\cdot\by_\bp=\bq\cdot\bz_\bq=0$,
we find the following expressions for the mode-to-mode energy transfer rates (see Appendix \ref{sec:derivation mtmt kinetic energy} and \ref{sec:derivation mtmt magnetic energy})
\begin{eqnarray}
T_E(\bu_\bp\overset{\bu_\bq}{\longrightarrow}\bu_\bk) =&&\Imag\{(\bp\cdot\bu_\bq)(\bu_\bp\cdot\bu_\bk)\}+\Delta_E^u(\bk|\bp|\bq),\label{Tuu}\\
T_E(\bb_\bp\overset{\bb_\bq}{\longrightarrow}\bu_\bk) =&-& \Imag\{(\bp\cdot\bb_\bq)(\bb_\bp\cdot\bu_\bk)\}+\Delta_E^b(\bp|\bq|\bk),\label{Tub}\\
T_E(\bb_\bp\overset{\bu_\bq}{\longrightarrow}\bb_\bk) =&&\Imag\{(\bp\cdot\bu_\bq)(\bb_\bp\cdot\bb_\bk)\}+\Delta_E^b(\bk|\bp|\bq) ,\label{Tbb}\\
T_E(\bu_\bp\overset{\bb_\bq}{\longrightarrow}\bb_\bk) =&-& \Imag\{(\bp\cdot\bb_\bq)(\bu_\bp\cdot\bb_\bk)\}+\Delta_E^b(\bq|\bk|\bp),\label{Tbu}
\end{eqnarray}
with 
\begin{eqnarray}
\Delta_E^u(\bk|\bp|\bq)&=&\alpha_E^u\Real\{(\bu_{\bk}, \bu_{\bp}, \bw_{\bq})
+ (\bu_{\bk}, \bw_{\bp}, \bu_{\bq})
+(\bw_{\bk}, \bu_{\bp}, \bu_{\bq})\}\label{eq:DeltaEu}\\
\Delta_E^b(\bk|\bp|\bq)&=&\alpha_E^b\Real\{(\bb_{\bk}, \bj_{\bp}, \bu_{\bq})
+ (\bj_{\bk}, \bb_{\bp}, \bu_{\bq})\}
+\beta_E^b\Real\{(\bb_{\bk}, \bb_{\bp}, \bw_{\bq})\},\label{eq:DeltaEb}
\end{eqnarray}
where $\alpha_E^u$, $\alpha_E^b$ and $\beta_E^b$
are real scalar quantities. The four types of energy transfer are represented in figure \ref{fig: Energy transfers}, with subscript $E$.

In (\ref{Tuu}-\ref{Tbu}) the first terms on the right hand sides coincide with the energy mode-to-mode transfer rates of \cite{DAR2001}. 
We note that in (\ref{Tuu}) $\Delta_E^u$ is invariant under cyclic permutation, $\Delta_E^u(\bk|\bp|\bq)=\Delta_E^u(\bq|\bk|\bp)=\Delta_E^u(\bp|\bq|\bk)$. Within the three modes $\bb_\bk, \bb_\bp$ and $\bu_\bq$ the additional terms in the transfer rates $\bb_\bp$-to-$\bb_\bk$, $\bb_\bk$-to-$\bu_\bq$ and $\bu_\bq$-to-$\bb_\bb$ are all equal to $\Delta_E^b(\bk|\bp|\bq)$. 
Therefore $\Delta_E^u$ in (\ref{Tuu}) and $\Delta_E^b$ in (\ref{Tub}-\ref{Tbu}) correspond to two circulating transfer rates in the sense that they do not change the kinetic and magnetic energy at each mode $\bk,\bp,\bq$. 
We note that $\Delta_E^u(\bk|\bp|\bq)=-\Delta_E^u(\bp|\bk|\bq)$ and $\Delta_E^b(\bk|\bp|\bq)=-\Delta_E^b(\bp|\bk|\bq)$. 

Replacing (\ref{Tbb}) and (\ref{Tbu}) in (\ref{eq:SversusT}) to calculate $S_E^{bb}(\bk|\bp,\bq)$ and $S_E^{bu}(\bk|\bp,\bq)$ leads to
\begin{eqnarray}
S_E^{bb}(\bk|\bp,\bq) =&&\Imag\{(\bp\cdot\bu_\bq)(\bb_\bp\cdot\bb_\bk)+(\bq\cdot\bu_\bp)(\bb_\bq\cdot\bb_\bk)\}
+\Delta_E^b(\bk|\bp|\bq)+\Delta_E^b(\bk|\bq|\bp),\label{Sbb} \;\;\;\;\;\;\\
S_E^{bu}(\bk|\bp,\bq) =&-&\Imag\{(\bp\cdot\bb_\bq)(\bu_\bp\cdot\bb_\bk)+(\bq\cdot\bb_\bp)(\bu_\bq\cdot\bb_\bk)\}
-\Delta_E^b(\bk|\bp|\bq)-\Delta_E^b(\bk|\bq|\bp). \;\;\;\;\;\;\label{Sbu}
\end{eqnarray}
In (\ref{Sbu}) we used the relations $\Delta_E^b(\bq|\bk|\bp)=-\Delta_E^b(\bk|\bq|\bp)$ and $\Delta_E^b(\bp|\bk|\bq)=-\Delta_E^b(\bk|\bp|\bq)$.
The two first terms on the right-hand sides of (\ref{Sbb}) and (\ref{Sbu}) coincide with the nonlinear energy transfer rates of \cite{DAR2001}.
The additional terms $\pm[\Delta_E^b(\bk|\bp|\bq)+\Delta_E^b(\bk|\bq|\bp)]$ may imply the non-uniqueness of the transfers $S_E^{bb}(\bk|\bp,\bq)$ and $S_E^{bu}(\bk|\bp,\bq)$, as anticipated from (\ref{SE3}).

Now let us write the induction equation (\ref{eq:b1}) in the following form
\begin{equation}
\partial_t \bb  +
\eta\nabla^2 {\bf b}  = - (\bu\cdot\nabla)\bb + (\bb\cdot\nabla)\bu,  \\
\end{equation}
and let us assume that the stretching term $(\bb\cdot\nabla)\bu$ and the advection term $ - (\bu\cdot\nabla)\bb$  correspond, in Fourier space, to respectively $S_E^{bu}\left(\bk|\bp,\bq\right)$ and $S_E^{bb}\left(\bk|\bp,\bq\right)$. 
Then the definitions of $S_E^{bu}\left(\bk|\bp,\bq\right)$ and $S_E^{bb}\left(\bk|\bp,\bq\right)$ become unique and correspond to (\ref{Sbb}) and (\ref{Sbu}) with
\begin{equation}
 \Delta_E^b(\bk|\bp|\bq)+\Delta_E^b(\bk|\bq|\bp)=0.
 \label{eq:sumdelta}
 \end{equation}
 On top of that, in (\ref{eq:sumdelta}) replacing $\Delta_E^b$ by its definition (\ref{eq:DeltaEb}) leads to $\alpha_E^b=\beta_E^b=0$, and then to $\Delta_E^b=0$. As a result the three MHD mode-to-mode transfers $ T_E(\bb_\bp\overset{\bb_\bq}{\longrightarrow}\bu_\bk)$, $T_E(\bb_\bp\overset{\bu_\bq}{\longrightarrow}\bb_\bk)$ and $T_E(\bu_\bp\overset{\bb_\bq}{\longrightarrow}\bb_\bk)$ then become uniquely defined. 
This is in striking contrast to the hydrodynamic case for which $T_E(\bu_\bp\overset{\bu_\bq}{\longrightarrow}\bu_\bk)$ is non-uniquely defined even though $S_E^{uu}\left(\bk|\bp,\bq\right)$ is uniquely defined. Thus Kraichnan's warning holds for the kinetic energy transfers in hydrodynamics and MHD, but can be withdrawn for the magnetic energy transfers in MHD, provided $S_E^{bu}$ and $S_E^{bb}$ correspond to respectively stretching and advection.

\subsection{Kinetic and magnetic helicity mode-to-mode transfer rates}
\label{sec:helmtm}
Following the same ideas as in section \ref{sec:enmtm}, after (\ref{SH1}) we define a basis of functions in real space $\Real\{(\bw_{\bk}, \bw_{\bp}, \bu_{\bq})\}, \Real\{(\bw_{\bk}, \bu_{\bp}, \bw_{\bq})\}$ and $\Real\{(\bu_{\bk}, \bw_{\bp}, \bw_{\bq})\}$, such that the kinetic helicity mode-to-mode transfer rate
$T_H(\bu_\bp\overset{\bu_\bq}{\longrightarrow}\bu_\bk)$ is expressed as a linear combination of them
\begin{eqnarray}
T_H(\bu_\bp\overset{\bu_\bq}{\longrightarrow}\bu_\bk) &=& A_4\Real\{(\bw_{\bk}, \bw_{\bp}, \bu_{\bq})\}
+ B_4\Real\{(\bw_{\bk}, \bu_{\bp}, \bw_{\bq})\}
+C_4\Real\{(\bu_{\bk}, \bw_{\bp}, \bw_{\bq})\},\;\;\;\;\;\;\;\;\;
\label{basisuuH}
\end{eqnarray}
where the coefficients $A_4,B_4, C_4$ are again scalar quantities.
Applying the two conditions (\ref{eq:SversusT}) and (\ref{eq:Topposite}) together with (\ref{SH1}) leads to the following expression (see Appendix \ref{sec:derivation mtmt kinetic helicity})
\begin{eqnarray}
T_H(\bu_\bp\overset{\bu_\bq}{\longrightarrow}\bu_\bk) &=& -\Real\{(\bw_\bk,\bw_\bp,\bu_\bq)\}+\Delta_H^u(\bk|\bp|\bq),\label{TuuH}
\end{eqnarray}
with 
\begin{eqnarray}
\Delta_H^u(\bk|\bp|\bq)&=&\alpha_H^u\Real\{(\bw_{\bk}, \bw_{\bp}, \bu_{\bq})
+ (\bw_{\bk}, \bu_{\bp}, \bw_{\bq})
+(\bu_{\bk}, \bw_{\bp}, \bw_{\bq})\}\label{eq:DeltaHu}
\end{eqnarray}
where $\alpha_H^u$
is a real scalar quantity. 
As the term $\Delta_H^u$ is invariant under cyclic permutation $\Delta_H^u(\bk|\bp|\bq)=\Delta_H^u(\bq|\bk|\bp)=\Delta_H^u(\bp|\bq|\bk)$ it corresponds again to some circulating quantity which does not change the helicity at each mode $\bk,\bp,\bq$. This shows that $T_H(\bu_\bp\overset{\bu_\bq}{\longrightarrow}\bu_\bk)$
is non-uniquely defined even though $S_H^{uu}\left(\bk|\bp,\bq\right)$ is uniquely defined.

The magnetic helicity transfer rate given in  (\ref{SH3})
depends on $\bu$ and $\bb$, but not on their derivatives $\bw$ and $\bj$. As a result,
from (\ref{SH3}), (\ref{eq:SversusT}) and (\ref{eq:Topposite}), the only choice for the mode-to-mode magnetic helicity transfer rate is 
\begin{eqnarray}
T_H(\bb_\bp\overset{\bu_\bq}{\longrightarrow}\bb_\bk) =-\Real\{(\bb_\bk,\bb_\bp,\bu_\bq)\},\label{TbbH}
\end{eqnarray}
which is then uniquely defined.

The two types of mode-to-mode helicity transfer rate, kinetic and magnetic, are represented in figure \ref{fig: Energy transfers}, with subscript $H$.

\section{Shell-to-shell transfer rates and fluxes}
\label{sec:conclusion}
Energy and helicity shell-to-shell transfer rates naturally derive from mode-to-mode transfer rates as
\begin{eqnarray}
\label{eq:tauE}
\tau_E(Y_P\rightarrow X_K)&=&\sum_{\bk\in K}\sum_{\bp\in P}T_E(\by_\bp\overset{\bz_\bq}{\longrightarrow}\bx_\bk)\\
\tau_H(X_P\rightarrow X_K)&=&\sum_{\bk\in K}\sum_{\bp\in P}T_H(\bx_\bp\overset{\bz_\bq}{\longrightarrow}\bx_\bk), \label{eq:tauH}
\end{eqnarray}
where $\tau_{E,H}(Y_P\rightarrow X_K)$ denotes the transfer rate of energy, or helicity (with $Y=X$), from $Y$ belonging to shell $P$ to $X$ belonging to shell $K$, $X$ and $Y$ being either $U$ or $B$. Typically it is the calculation of these quantities which allows us to determine the degree of locality of the energy or helicity transfers \citep{Alexakis2005,Mininni2005,Mininni2011}. Mapping the value of 
$\tau_{E,H}(Y_P\rightarrow X_K)$ versus shells $P$ and $K$,
if the maxima are reached for neighbouring shells then the transfers are local. On the other hand if the maxima are reached for distant shells then the transfers are non-local.

After ($\ref{eq:tauE}$) and ($\ref{eq:tauH}$) there is equivalence of uniqueness between shell-to-shell transfer rates and mode-to-mode transfer rates. Therefore the uniquely defined shell-to-shell transfer rate is $\tau_{H}(B_P\rightarrow B_K)$, but also $\tau_{E}(B_P\rightarrow U_K)$, $\tau_{E}(B_P\rightarrow B_K)$ and $\tau_{E}(U_P\rightarrow B_K)$ provided $S_E^{bu}$ and $S_E^{bb}$ correspond to respectively stretching and advection (see section \ref{sec:enmtm}). In contrast the shell-to-shell transfer rate $\tau_{E,H}(U_P\rightarrow U_K)$ is not uniquely defined.

The lack of unicity of the kinetic mode-to-mode and shell-to-shell transfers render their numerical calculations rather hazardous. Indeed, we found that the mode-to-mode transfer rates $T_{E,H}(\bu_\bp\overset{\bu_\bq}{\longrightarrow}\bu_\bk)$ depend on the circulating terms 
$\Delta_{E,H}^u$ which in general are non-zero and depend on scalar coefficients $\alpha_{E,H}^u$ that can be chosen arbitrarily. In addition, these coefficients may differ, again arbitrarily, from one triad to another.Thus, for a given set of velocity fields one could imagine adjusting these free coefficients to find, on demand, either local or non-local shell-to-shell transfer rates. 
Of course, this is not satisfactory as the results should depend on the physics and not on arbitrary choices.

Now, let us denote by $\Pi^{x<}_{y>}(k_0)$ the flux of energy from $\bx_{\bp, p\le k_0}$ to $\by_{\bk, k>k_0}$. It corresponds to a shell-to-shell transfer between two adjacent shells,
 $P$ being the inner shell and $K$ the outer shell, with $P$ extending to the minimum wavenumber and $K$ to the maximum one, such that $\bq$ necessarily belongs either to $P$ or $K$.
The frontier between $P$ and $K$ corresponds to $k=k_0$, $\tau_{E}(Y_P\rightarrow X_K)$ coinciding with $\Pi^{y<}_{x>}(k_0)$. The latter is then defined by
\begin{equation}
\Pi^{y<}_{x>}(k_0)=\sum_{p\le k_0}\sum_{k>k_0}T_E(\by_\bp\overset{\bz_\bq}{\longrightarrow}\bx_\bk). \label{eq:flux}
\end{equation}
Similarly, the following fluxes are defined by: 
\begin{eqnarray}
\Pi^{y<}_{x<}(k_0)&=&\sum_{p\le k_0}\sum_{k\le k_0}T_E(\by_\bp\overset{\bz_\bq}{\longrightarrow}\bx_\bk), \label{eq:flux2}\\
\Pi^{y>}_{x>}(k_0)&=&\sum_{p> k_0}\sum_{k> k_0}T_E(\by_\bp\overset{\bz_\bq}{\longrightarrow}\bx_\bk),\label{eq:flux3}\\
\Pi^{y>}_{x<}(k_0)&=&-\Pi^{x<}_{y>}(k_0).  \label{eq:flux4}
\end{eqnarray}
Strictly speaking, these quantities correspond to fluxes only if $\bx=\by=\bu$.  In the other cases it is an abuse of notation, in that they are not associated with any conserved quantity. To stress the difference the latter are denoted with quotation marks ``fluxes''.

Splitting the right hand side of (\ref{eq:flux}) as the sum of two terms for $q\le k_0$ and $q> k_0$,
it is shown in Appendix \ref{sec:derivation Fluxes} that 
\begin{eqnarray}
\Pi^{y<}_{x>}(k_0)&=&\frac{1}{2}\sum \limits_{\underset{p,q\le k_0}{\bp,\bq}}\sum \limits_{\underset{k> k_0}{\bk}}S^{xy}_E(\bk|\bp,\bq) 
- \frac{1}{2}\sum \limits_{\underset{p,q> k_0}{\bp,\bq}}\sum \limits_{\underset{k\le k_0}{\bk}}S^{yx}_E(\bk|\bp,\bq), \label{eq:flux6}
\end{eqnarray}
leading to
\begin{eqnarray}
\Pi^{u<}_{u>}(k_0)&=&\frac{1}{2}\sum \limits_{\underset{p,q\le k_0}{\bp,\bq}}\sum \limits_{\underset{k> k_0}{\bk}}S^{uu}_E(\bk|\bp,\bq)
- \frac{1}{2}\sum \limits_{\underset{p,q> k_0}{\bp,\bq}}\sum \limits_{\underset{k\le k_0}{\bk}}S^{uu}_E(\bk|\bp,\bq)\label{eq:fluxuu}\\
\Pi^{b<}_{u>}(k_0)&=&\frac{1}{2}\sum \limits_{\underset{p,q\le k_0}{\bp,\bq}}\sum\limits_{\underset{k> k_0}{\bk}}S^{ub}_E(\bk|\bp,\bq)
- \frac{1}{2}\sum \limits_{\underset{p,q> k_0}{\bp,\bq}}\sum \limits_{\underset{k\le k_0}{\bk}}S^{bu}_E(\bk|\bp,\bq), \label{eq:fluxbu}\\
\Pi^{b<}_{b>}(k_0)&=&\frac{1}{2}\sum \limits_{\underset{p,q\le k_0}{\bp,\bq}}\sum \limits_{\underset{k> k_0}{\bk}}S^{bb}_E(\bk|\bp,\bq)
- \frac{1}{2}\sum \limits_{\underset{p,q> k_0}{\bp,\bq}}\sum \limits_{\underset{k\le k_0}{\bk}}S^{bb}_E(\bk|\bp,\bq), \label{eq:fluxbb}\\
\Pi^{u<}_{b>}(k_0)&=&\frac{1}{2}\sum \limits_{\underset{p,q\le k_0}{\bp,\bq}}\sum \limits_{\underset{k> k_0}{\bk}}S^{bu}_E(\bk|\bp,\bq)
- \frac{1}{2}\sum \limits_{\underset{p,q> k_0}{\bp,\bq}}\sum \limits_{\underset{k\le k_0}{\bk}}S^{ub}_E(\bk|\bp,\bq),\label{eq:fluxub}
\end{eqnarray}
where $\sum \limits_{\underset{p,q\le k_0}{\bp,\bq}}=\sum \limits_{\underset{p\le k_0}{\bp}}\sum \limits_{\underset{q\le k_0}{\bq}}$,
$\sum \limits_{\underset{p,q> k_0}{\bp,\bq}}=\sum \limits_{\underset{p> k_0}{\bp}}\sum \limits_{\underset{q> k_0}{\bq}}$, and $\bk+\bp+\bq=0$.

From (\ref{eq:fluxuu}), which was first introduced by \cite{kraichnan1959},  we immediately see that, as $S_{E}^{uu}(\bk|\bp,\bq)$ is uniquely defined, so is the flux of kinetic energy $\Pi^{u<}_{u>}(k_0)$.
The three other ``fluxes'' $\Pi^{b<}_{u>}(k_0)$, $\Pi^{b<}_{b>}(k_0)$ and $\Pi^{u<}_{b>}(k_0)$ are uniquely defined provided again that $S_E^{bu}$ and $S_E^{bb}$ correspond to respectively stretching and advection. 

\begin{figure*}
\centering
\includegraphics[width=0.5\columnwidth]{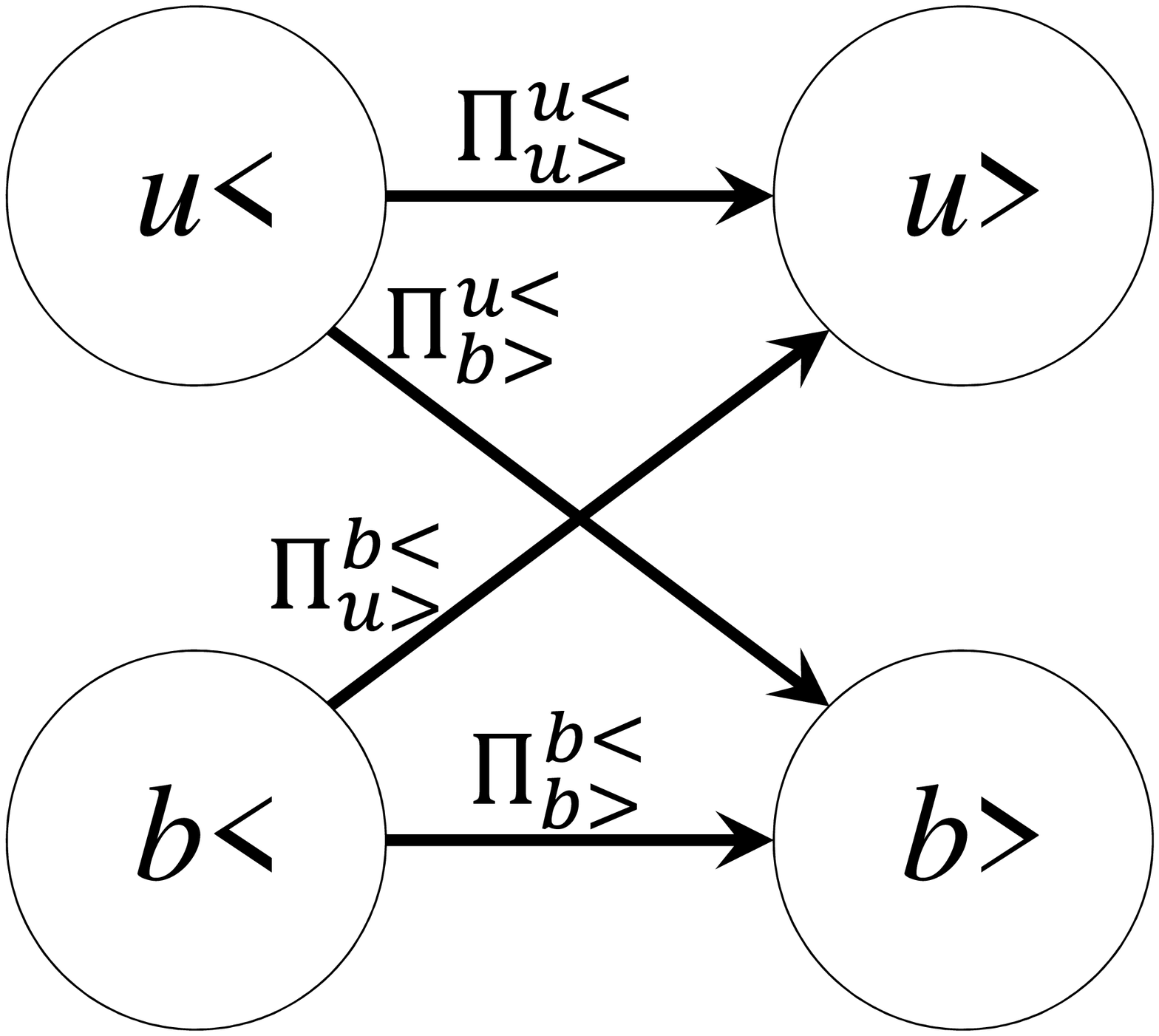}\hfill
\includegraphics[width=0.5\columnwidth]{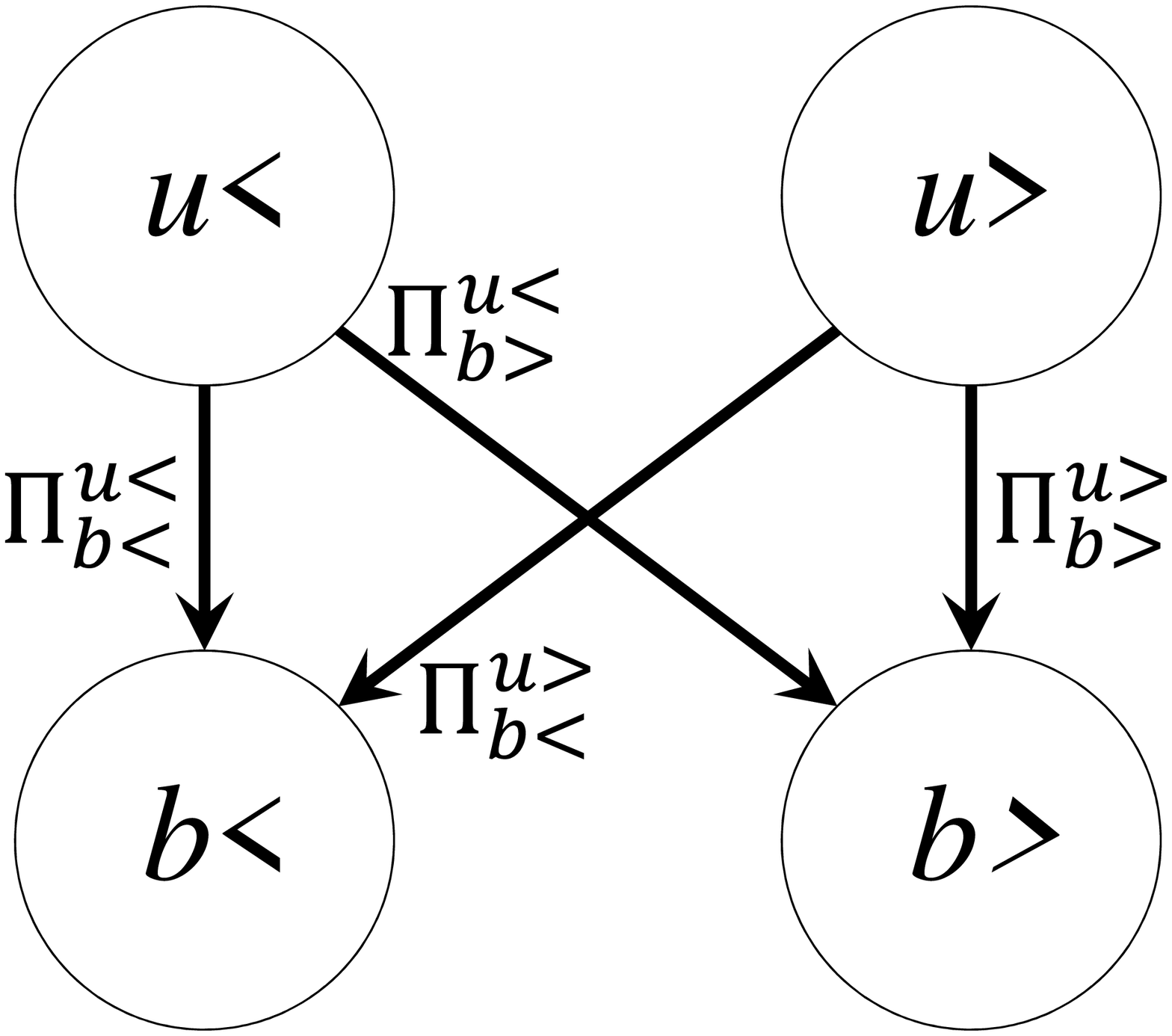}
\caption{\label{fig: Energy fluxes} The sphere of radius $k_0$ separates the Fourier space in two parts the inner space $|\bk|\le k_0$ denoted by $<$ and the outer space $|\bk|> k_0$ denoted by $>$. 
On the left, the arrows describe the ``fluxes'' whose sum is the total flux of energy at $k=k_0$. On the right, the arrows describe the ``fluxes''   whose sum is independant on $k_0$.}
\end{figure*}

In addition (\ref{eq:fluxbu}-\ref{eq:fluxub}) lead to 
\begin{eqnarray}
\Pi^{b<}_{u>}(k_0)+\Pi^{b<}_{b>}(k_0)+\Pi^{u<}_{b>}(k_0)=
\frac{1}{2}\sum \limits_{\underset{p,q\le k_0}{\bp,\bq}}\sum\limits_{\underset{k> k_0}{\bk}}(S^{ub}_E+S^{bb}_E+S^{bu}_E)(\bk|\bp,\bq) \nonumber \\
- \frac{1}{2}\sum \limits_{\underset{p,q> k_0}{\bp,\bq}}\sum\limits_{\underset{k\le k_0}{\bk}}(S^{ub}_E+S^{bb}_E+S^{bu}_E)(\bk|\bp,\bq),\label{eq:sumflux}
\end{eqnarray}
finally implying with (\ref{eq:fluxuu}) that the flux of total energy is always uniquely defined.
Besides, it is straightforward to show that the sum on all $k_0$ of respectively (\ref{eq:fluxuu}) and (\ref{eq:sumflux}) is equal to zero.
An illustration of the ``fluxes'' involved in the total flux of energy is given on Figure \ref{fig: Energy fluxes}(a).

There is also a remarkable identity
\begin{equation}
\frac{\partial}{ \partial k_0}\left(\Pi^{u<}_{b<}(k_0)+\Pi^{u<}_{b>}(k_0)+\Pi^{u>}_{b<}(k_0)+\Pi^{u>}_{b>}(k_0)\right)=0
\label{eq:sumfluxcste}
\end{equation}
which is exactly satisfied whatever the value of $k_0$, including in the infrared and dissipation ranges, and whether the turbulence state is stationary or not. To demonstrate this, we just replace in (\ref{eq:sumfluxcste}) the ``fluxes'' by their definitions (\ref{eq:flux}-\ref{eq:flux4}), leading to
\begin{equation}
\left(
   \sum \limits_{\underset{p,k\le k_0}{\bp,\bk}}
+ \sum \limits_{\underset{p\le k_0, k>k_0}{\bp,\bk}}
+ \sum \limits_{\underset{p> k_0, k\le k_0}{\bp,\bk}}
+ \sum \limits_{\underset{p, k>k_0}{\bp,\bk}}
\right)
T_E(\bu_\bp\overset{\bb_\bq}{\longrightarrow}\bb_\bk)
= \sum_{\bp}\sum_{\bk}
T_E(\bu_\bp\overset{\bb_\bq}{\longrightarrow}\bb_\bk),
\label{eq:sumindk0}
\end{equation}
which does not depend on $k_0$.
This can also be shown from the magnetic energy equation
(\ref{Energy_Fourier_timeb2}) written in the following form
\begin{equation}
\label{eq:EbfromT}
( \partial_t  + 2\eta k^2) E^b_\bk  = \sum \limits_{\underset{\bf k+p+q=0}{\bp}}
T_E(\bu_\bp\overset{\bb_\bq}{\longrightarrow}\bb_\bk)+T_E(\bb_\bp\overset{\bu_\bq}{\longrightarrow}\bb_\bk),
\end{equation}
by taking the sum over all $\bk$, and noticing that $\Pi^{x<}_{x<}(k_0)=\Pi^{x>}_{x>}(k_0)=0$ (See Appendix \ref{App:Flux Identity}). An illustration of the ``fluxes'' involved in (\ref{eq:sumfluxcste}) is given on Figure \ref{fig: Energy fluxes}(b).

Finally, the kinetic and magnetic helicity fluxes are given by equations (\ref{eq:fluxuu}) and (\ref{eq:fluxbb}) in which the subscript $E$ is replaced by $H$. They are uniquely defined, consistent with the helicity conservation laws, equation (\ref{SumSuuuhel}) in hydrodynamic and  (\ref{SumShelb}) in MHD.

\section{Conclusion}
In this paper, using appropriate basis of functions, we derived the complete expressions for the energy and helicity mode-to-mode transfer rates for the MHD equations. With the exception of magnetic helicity, all other mode-to-mode transfer rate definitions  include additional terms corresponding to circulating transfers, each of them depending on one or two arbitrary scalar quantities.  
Although these circulating transfers do not change the energy and helicity in each mode, in principle they can render the mode-to-mode transfer rates non-uniquely defined. For shell-to-shell transfer rates the conclusion is the same.

However, in the induction equation we can always split the nonlinear term into a magnetic stretching term and a magnetic advection term.
Then it may be physically relevant, in Fourier space, to identify the energy transfer rates $S_E^{bu}$ ($\bu$-to-$\bb$) and $S_E^{bb}$ ($\bb$-to-$\bb$) as coming from respectively   the magnetic stretching and the magnetic advection. In that case we find that the circulating terms in the expression for the magnetic energy mode-to-mode transfer rates are equal to zero, implying unique definitions of
$S_E^{bu}$ and  $S_E^{bb}$. Then all magnetic ($\bu$-to-$\bb$, $\bb$-to-$\bb$ and $\bb$-to-$\bu$) mode-to-mode and shell-to-shell transfer rates also become uniquely defined. 
Only the kinetic ($\bu$-to-$\bu$) energy and helicity mode-to-mode and shell-to-shell transfer rates remain non-uniquely defined, as already stressed by \cite{kraichnan1959} for the energy transfer rate.
Such a difference between the hydrodynamic and MHD cases is probably a consequence of the fact that, in MHD, transfer rates always occur between two modes of one field (magnetic) and one  mode of another field (velocity),  in contrast to the hydrodynamic case involving three modes of the same field (velocity), as shown in figure \ref{fig: MHD triad}.

Applying the flux definition (\ref{eq:flux}) to energy and helicity (then replacing the subscript $E$ by $H$) we find that the fluxes of kinetic energy, kinetic helicity, magnetic helicity and total energy are uniquely defined. This is also a direct consequence of the conservation laws given in section \ref{sec:generalMHD}. Provided again that $S_E^{bu}$ and $S_E^{bb}$ correspond to respectively magnetic stretching and magnetic advection, 
the three other MHD ``fluxes'' of energy $\bb$-to-$\bu$,  $\bb$-to-$\bb$ and $\bu$-to-$\bb$ are also uniquely defined.
 
In summary,  Kraichnan's original warning about the lack of unicity of mode-to-mode (and shell-to-shell) transfers in hydrodynamics does not apply to the magnetic transfers in MHD for magnetic helicity. Nor does it apply to magnetic energy, provided that magnetic stretching and advection are identified as energy transfer channels.

This work was supported by the Department of Science and Technology, India (INT/RUS/RSF/P-03). 
R.S. acknowledges support from the Russian Foundation for
Basic Research and the Perm regional government under project RFBR-17-41-
590059.
F.P. also thanks support from IFCAM project MA/IFCAM/19/90, and IITK for hospitality.

\bibliographystyle{jpp}
\bibliography{refs}

\appendix
\section{Derivation of identities (\ref{identity1}) and (\ref{identity2})}
\label{Appsec:identity}
Here we demonstrate the identity (\ref{identity2}), the identity  (\ref{identity1}) resulting from  (\ref{identity2}) by replacing $\bx, \by$ and $\bz$ by $\bu$.

We have
\begin{eqnarray} 
 \left(\bx_{\bk}, \by_{\bp}, \ii \bq \times\bz_{\bq}\right)&=&\bx_{\bk} \cdot[\by_{\bp} \times (\ii \bq \times \bz_{\bq})]\\
 &=&\bx_{\bk} \cdot[(\by_{\bp}\cdot \bz_{\bq})\ii \bq -\ii (\bq \cdot \by_{\bp}) \bz_{\bq}]\\
 &=&\ii (\bx_{\bk} \cdot\bq) (\by_{\bp}\cdot \bz_{\bq})  -\ii (\bq \cdot \by_{\bp}) (\bz_{\bq}\cdot \bx_{\bk})\label{App:identity1}
\end{eqnarray}
After a similar type of algebra we find
\begin{eqnarray} \label{App:identity2}
- \left(\bx_{\bk}, \ii \bp \times\by_{\bp}, \bz_{\bq}\right)
 &=&\ii (\bx_{\bk} \cdot\bp) (\by_{\bp}\cdot \bz_{\bq})  -\ii (\bp \cdot \bz_{\bq}) (\by_{\bp}\cdot \bx_{\bk})
\end{eqnarray}
and
\begin{eqnarray} \label{App:identity3}
- \left(\ii \bk \times\bx_{\bk}, \by_{\bp}, \bz_{\bq}\right)
 &=&\ii (\bz_{\bq} \cdot\bk) (\by_{\bp}\cdot \bx_{\bk})  -\ii (\bk \cdot \by_{\bp}) (\bz_{\bq}\cdot \bx_{\bk}).
\end{eqnarray}
Adding (\ref{App:identity1}), (\ref{App:identity2}) and (\ref{App:identity3}) leads to
\begin{eqnarray} \label{App:identity4}
&&\left(\bx_{\bk}, \by_{\bp}, \ii \bq \times\bz_{\bq}\right)- \left(\bx_{\bk}, \ii\bp\times\by_{\bp}, \bz_{\bq}\right)  - \left(\ii\bk\times\bx_{\bk}, \by_{\bp}, \bz_{\bq}\right)\nonumber\\
&=&\ii [\bx_{\bk} \cdot(\bq+\bp)] (\by_{\bp}\cdot \bz_{\bq}) - \ii [ (\bq+ \bk) \cdot \by_{\bp}] (\bz_{\bq}\cdot \bx_{\bk})+\ii [\bz_{\bq} \cdot(\bk-\bp)] (\by_{\bp}\cdot \bx_{\bk}).
\end{eqnarray}
Applying the triadic condition $\bk+\bp+\bq=0$ and the solenoidality conditions $\bk\cdot\bx_\bk=\bp\cdot\by_\bp=\bq\cdot\bz_\bq=0$, leads to
\begin{eqnarray} \label{App:identity5}
\left(\bx_{\bk}, \by_{\bp}, \ii \bq \times\bz_{\bq}\right)- \left(\bx_{\bk}, \ii\bp\times\by_{\bp}, \bz_{\bq}\right)  - \left(\ii\bk\times\bx_{\bk}, \by_{\bp}, \bz_{\bq}\right)
&=&-2\ii (\bz_{\bq} \cdot\bp) (\by_{\bp}\cdot \bx_{\bk}).
\end{eqnarray}
which is (\ref{identity2}).
\section{Derivation of the mode-to-mode transfers}
\label{sec:derivation mtmt}
\subsection{Derivation of kinetic energy mode-to-mode transfer rate (\ref{Tuu})}
\label{sec:derivation mtmt kinetic energy}
From (\ref{eq:SversusT}) we have
\begin{eqnarray}
S_E^{uu}(\bk|\bp,\bq) &=& T_E(\bu_\bp\overset{\bu_\bq}{\longrightarrow}\bu_\bk)+T_E(\bu_\bq\overset{\bu_\bp}{\longrightarrow}\bu_\bk), \label{App:SuversusTu}
\end{eqnarray}
with $S_E^{uu}$ defined in (\ref{SE1}) as
\begin{equation}
S_E^{uu}\left(\bk|\bp,\bq\right)=\Real\{(\bu_\bk,\bu_\bp,\bw_\bq)-(\bu_\bk,\bw_\bp,\bu_\bq)\},\label{App:SE1}\\
\end{equation}
and $T_E(\bu_\bp\overset{\bu_\bq}{\longrightarrow}\bu_\bk)$ and $T_E(\bu_\bq\overset{\bu_\bp}{\longrightarrow}\bu_\bk)$ defined from (\ref{basisuu}) as
\begin{eqnarray}
T_E(\bu_\bp\overset{\bu_\bq}{\longrightarrow}\bu_\bk) &=& A_0\Real\{(\bu_{\bk}, \bu_{\bp}, \bw_{\bq})\}
+ B_0\Real\{(\bu_{\bk}, \bw_{\bp}, \bu_{\bq})\}
+C_0\Real\{(\bw_{\bk}, \bu_{\bp}, \bu_{\bq})\},\;\;\;\;\;\;\label{App:basisuu1}\\
T_E(\bu_\bq\overset{\bu_\bp}{\longrightarrow}\bu_\bk) &=& A_0\Real\{(\bu_{\bk}, \bu_{\bq}, \bw_{\bp})\}
+ B_0\Real\{(\bu_{\bk}, \bw_{\bq}, \bu_{\bp})\}
+C_0\Real\{(\bw_{\bk}, \bu_{\bq}, \bu_{\bp})\},\;\;\;\;\;\;\label{App:basisuu2}
\end{eqnarray}
where (\ref{App:basisuu2}) is obtained from (\ref{App:basisuu1}) by exchanging $\bp$ and $\bq$.
Replacing (\ref{App:SE1}), (\ref{App:basisuu1}) and (\ref{App:basisuu2}) in (\ref{App:SuversusTu}) leads to $A_0-B_0=1$.

From (\ref{eq:Topposite}) we have
\begin{eqnarray}
\label{App:Topposite}
T_{E}(\bu_\bp\overset{\bu_\bq}{\longrightarrow}\bu_\bk)&=&
-T_{E}(\bu_\bk\overset{\bu_\bq}{\longrightarrow}\bu_\bp).
\end{eqnarray}
with
\begin{eqnarray}
T_E(\bu_\bk\overset{\bu_\bq}{\longrightarrow}\bu_\bp) &=& A_0\Real\{(\bu_{\bp}, \bu_{\bk}, \bw_{\bq})\}
+ B_0\Real\{(\bu_{\bp}, \bw_{\bk}, \bu_{\bq})\}
+C_0\Real\{(\bw_{\bp}, \bu_{\bk}, \bu_{\bq})\},\;\;\;\;\;\;\label{App:basisuu3}
\end{eqnarray}
which has been obtained from (\ref{App:basisuu1}) by exchanging $\bp$ and $\bk$.
Replacing (\ref{App:basisuu1}) and (\ref{App:basisuu3}) in (\ref{App:Topposite}) leads to $B_0=C_0$.
Taking $A_0=\frac{1}{2}+\alpha_E^u$, $B_0=C_0=-\frac{1}{2}+\alpha_E^u$ in (\ref{App:basisuu1})  leads to 
\begin{eqnarray}
T_E(\bu_\bp\overset{\bu_\bq}{\longrightarrow}\bu_\bk) &=& \frac{1}{2}\Real\{(\bu_{\bk}, \bu_{\bp}, \bw_{\bq})
-(\bu_{\bk}, \bw_{\bp}, \bu_{\bq})
-(\bw_{\bk}, \bu_{\bp}, \bu_{\bq})\}\nonumber\\
&+&\alpha_E^u \Real\{(\bu_{\bk}, \bu_{\bp}, \bw_{\bq})
+(\bu_{\bk}, \bw_{\bp}, \bu_{\bq})
+(\bw_{\bk}, \bu_{\bp}, \bu_{\bq})\}.\label{App:basisuu4}
\end{eqnarray}
In (\ref{identity2}) replacing $\bx, \by$ and $\bz$ by $\bu$, we have
\begin{equation} \label{App:identity2u}
 \left(\bu_{\bk}, \bw_{\bp}, \bu_{\bq}\right)  + \left(\bw_{\bk}, \bu_{\bp}, \bu_{\bq}\right)=\left(\bu_{\bk}, \bu_{\bp}, \bw_{\bq}\right)+ 2\ii(\bp\cdot\bu_{\bq})( \bu_{\bp}\cdot\bu_{\bk}).
\end{equation}
Then replacing (\ref{App:identity2u}) in (\ref{App:basisuu4}) leads to
\begin{eqnarray}
T_E(\bu_\bp\overset{\bu_\bq}{\longrightarrow}\bu_\bk) &=&\Imag\{(\bp\cdot\bu_{\bq})( \bu_{\bp}\cdot\bu_{\bk})\} +\Delta_E^u(\bk|\bp|\bq),\label{App:basisuu5}
\end{eqnarray}
with
\begin{equation}
\Delta_E^u(\bk|\bp|\bq)=\alpha_E^u\Real\{(\bu_{\bk}, \bu_{\bp}, \bw_{\bq})
+ (\bu_{\bk}, \bw_{\bp}, \bu_{\bq})
+(\bw_{\bk}, \bu_{\bp}, \bu_{\bq})\},\label{App:DeltaEu}
\end{equation}
which is (\ref{Tuu}) with (\ref{eq:DeltaEu}).

\subsection{Derivation of magnetic energy mode-to-mode transfer rates (\ref{Tub}), (\ref{Tbb}), (\ref{Tbu})}
\label{sec:derivation mtmt magnetic energy}
From (\ref{eq:SversusT}) we have
\begin{eqnarray}
S_E^{ub}(\bk|\bp,\bq) &=& T_E(\bb_\bp\overset{\bb_\bq}{\longrightarrow}\bu_\bk)+T_E(\bb_\bq\overset{\bb_\bp}{\longrightarrow}\bu_\bk), \label{App:SubversusTub}\\
S_E^{bb}(\bk|\bp,\bq) &=& T_E(\bb_\bp\overset{\bu_\bq}{\longrightarrow}\bb_\bk)+T_E(\bb_\bq\overset{\bu_\bp}{\longrightarrow}\bb_\bk), \label{App:SubversusTbb}\\
S_E^{bu}(\bk|\bp,\bq) &=& T_E(\bu_\bp\overset{\bb_\bq}{\longrightarrow}\bb_\bk)+T_E(\bu_\bq\overset{\bb_\bp}{\longrightarrow}\bb_\bk), \label{App:SubversusTbu}
\end{eqnarray}
with $S_E^{ub}, S_E^{bb}$ and $S_E^{bb}$ satisfying (\ref{SE2}) and (\ref{SE3})
\begin{eqnarray}
S_E^{ub}\left(\bk|\bp,\bq\right)=&\Real\{(\bu_\bk,\bj_\bp,\bb_\bq)&-(\bu_\bk,\bb_\bp,\bj_\bq)\},\label{App:SE2}\\
S_E^{bu}\left(\bk|\bp,\bq\right)+S_E^{bb}\left(\bk|\bp,\bq\right)=&\Real\{(\bj_\bk,\bu_\bp,\bb_\bq)&-(\bj_\bk,\bb_\bp,\bu_\bq)\},\label{App:SE3}
\end{eqnarray}
and,  from (\ref{basisub}-\ref{basisbu}),
\begin{eqnarray}
T_E(\bb_\bp\overset{\bb_\bq}{\longrightarrow}\bu_\bk) &=& A_1\Real\{(\bu_{\bk}, \bb_{\bp}, \bj_{\bq})\}
+ B_1\Real\{(\bu_{\bk}, \bj_{\bp}, \bb_{\bq})\}
+C_1\Real\{(\bw_{\bk}, \bb_{\bp}, \bb_{\bq})\},\;\;\;\;\;\;\label{App:basisub}\\
T_E(\bb_\bq\overset{\bb_\bp}{\longrightarrow}\bu_\bk) &=& A_1\Real\{(\bu_{\bk}, \bb_{\bq}, \bj_{\bp})\}
+ B_1\Real\{(\bu_{\bk}, \bj_{\bq}, \bb_{\bp})\}
+C_1\Real\{(\bw_{\bk}, \bb_{\bq}, \bb_{\bp})\},\label{App:basisub2}\\
T_E(\bb_\bp\overset{\bu_\bq}{\longrightarrow}\bb_\bk) &=& A_2\Real\{(\bb_{\bk}, \bb_{\bp}, \bw_{\bq})\}
+ B_2\Real\{(\bb_{\bk}, \bj_{\bp}, \bu_{\bq})\}
+C_2\Real\{(\bj_{\bk}, \bb_{\bp}, \bu_{\bq})\},\label{App:basisbb}\\
T_E(\bb_\bq\overset{\bu_\bp}{\longrightarrow}\bb_\bk) &=& A_2\Real\{(\bb_{\bk}, \bb_{\bq}, \bw_{\bp})\}
+ B_2\Real\{(\bb_{\bk}, \bj_{\bq}, \bu_{\bp})\}
+C_2\Real\{(\bj_{\bk}, \bb_{\bq}, \bu_{\bp})\},\label{App:basisb2}\\
T_E(\bu_\bp\overset{\bb_\bq}{\longrightarrow}\bb_\bk) &=& A_3\Real\{(\bb_{\bk}, \bu_{\bp}, \bj_{\bq})\}
+ B_3\Real\{(\bb_{\bk}, \bw_{\bp}, \bb_{\bq})\}
+C_3\Real\{(\bj_{\bk}, \bu_{\bp}, \bb_{\bq})\},\label{App:basisbu}\\
T_E(\bu_\bq\overset{\bb_\bp}{\longrightarrow}\bb_\bk) &=& A_3\Real\{(\bb_{\bk}, \bu_{\bq}, \bj_{\bp})\}
+ B_3\Real\{(\bb_{\bk}, \bw_{\bq}, \bb_{\bp})\}
+C_3\Real\{(\bj_{\bk}, \bu_{\bq}, \bb_{\bp})\}.\label{App:basisbu2}
\end{eqnarray}
Replacing (\ref{App:SE2}), (\ref{App:basisub}) and (\ref{App:basisub2}) in (\ref{App:SubversusTub}) leads to $B_1-A_1=1$.
Replacing (\ref{App:basisbb}-\ref{App:basisbu2}) in (\ref{App:SubversusTbb}) and  (\ref{App:SubversusTbu}), then (\ref{App:SubversusTbb}) and  (\ref{App:SubversusTbu}) in (\ref{App:SE3}), leads to $A_3=B_2, B_3=A_2$ and $C_3=1+C_2$.

From (\ref{eq:Topposite}) we have
\begin{eqnarray}
\label{App:Toppositebu}
T_{E}(\bb_\bp\overset{\bb_\bq}{\longrightarrow}\bu_\bk)&=&
-T_{E}(\bu_\bk\overset{\bb_\bq}{\longrightarrow}\bb_\bp),\\\label{App:Toppositebb}
T_{E}(\bb_\bp\overset{\bu_\bq}{\longrightarrow}\bb_\bk)&=&
-T_{E}(\bb_\bk\overset{\bu_\bq}{\longrightarrow}\bb_\bp),
\end{eqnarray}
with
\begin{eqnarray}
T_E(\bu_\bk\overset{\bb_\bq}{\longrightarrow}\bb_\bp) &=& A_3\Real\{(\bb_{\bp}, \bu_{\bk}, \bj_{\bq})\}
+ B_3\Real\{(\bb_{\bp}, \bw_{\bk}, \bb_{\bq})\}
+C_3\Real\{(\bj_{\bp}, \bu_{\bk}, \bb_{\bq})\},\;\;\;\;\;\;\label{App:basisbu3}\\
T_E(\bb_\bk\overset{\bu_\bq}{\longrightarrow}\bb_\bp) &=& A_2\Real\{(\bb_{\bp}, \bb_{\bk}, \bw_{\bq})\}
+ B_2\Real\{(\bb_{\bp}, \bj_{\bk}, \bu_{\bq})\}
+C_2\Real\{(\bj_{\bp}, \bb_{\bk}, \bu_{\bq})\}.\label{App:basisbb3}
\end{eqnarray}
Replacing (\ref{App:basisub}) and (\ref{App:basisbu3}) in (\ref{App:Toppositebu}) leads to $A_1=A_3, C_1=B_3$ and $B_1=C_3$.
Replacing (\ref{App:basisbb}) and (\ref{App:basisbb3}) in (\ref{App:Toppositebb}) leads to $B_2=C_2$. We can express all coefficients in terms of $A_1$ and $C_1$,
$B_1=C_3=1+A_1, B_2=C_2=A_3=A_1$ and $A_2=B_3=C_1$. 

Taking $A_1=\alpha_E^b-\frac{1}{2}$, $B_1=1+A_1$ and $C_1=\beta_E^b+\frac{1}{2}$ in (\ref{App:basisub})
leads to
\begin{eqnarray}
T_E(\bb_\bp\overset{\bb_\bq}{\longrightarrow}\bu_\bk) &=& \frac{1}{2}\Real\{-(\bu_{\bk}, \bb_{\bp}, \bj_{\bq})+(\bu_{\bk}, \bj_{\bp}, \bb_{\bq})+(\bw_{\bk}, \bb_{\bp}, \bb_{\bq})\} \nonumber\\
 &+& \alpha_E^b\Real\{(\bu_{\bk}, \bb_{\bp}, \bj_{\bq})+(\bu_{\bk}, \bj_{\bp}, \bb_{\bq})\}
+\beta_E^b\Real\{(\bw_{\bk}, \bb_{\bp}, \bb_{\bq})\}. \label{App:basisub4}
\end{eqnarray}
In (\ref{identity2}) replacing $\bx, \by$ and $\bz$ by respectively $\bu$, $\bb$ and $\bb$ we have
\begin{equation} \label{App:identity2b}
 \left(\bu_{\bk}, \bj_{\bp}, \bb_{\bq}\right)  + \left(\bw_{\bk}, \bb_{\bp}, \bb_{\bq}\right)=\left(\bu_{\bk}, \bb_{\bp}, \bj_{\bq}\right)+ 2\ii(\bp\cdot\bb_{\bq})( \bb_{\bp}\cdot\bu_{\bk}).
\end{equation}
Then replacing (\ref{App:identity2b}) in (\ref{App:basisub4}) leads to
\begin{eqnarray}
T_E(\bb_\bp\overset{\bb_\bq}{\longrightarrow}\bu_\bk) &=&-\Imag\{(\bp\cdot\bb_{\bq})( \bb_{\bp}\cdot\bu_{\bk})\} +\Delta_E^b(\bp|\bq|\bk),\label{App:basisub5}
\end{eqnarray}
with
\begin{equation}
\Delta_E^b(\bp|\bq|\bk)=\alpha_E^b\Real\{(\bu_{\bk}, \bb_{\bp}, \bj_{\bq})+(\bu_{\bk}, \bj_{\bp}, \bb_{\bq})\}
+\beta_E^b\Real\{(\bw_{\bk}, \bb_{\bp}, \bb_{\bq})\},\label{App:DeltaEb}
\end{equation}
which is (\ref{Tub}) with (\ref{eq:DeltaEb}).

In (\ref{App:basisbb}) replacing $A_2$ by $C_1=\beta_E^b+\frac{1}{2}$, $B_2$ and $C_2$ by $A_1=\alpha_E^b-\frac{1}{2}$ leads to
\begin{eqnarray}
T_E(\bb_\bp\overset{\bu_\bq}{\longrightarrow}\bb_\bk) &=& \frac{1}{2}\Real\{(\bb_{\bk}, \bb_{\bp}, \bw_{\bq})-(\bb_{\bk}, \bj_{\bp}, \bu_{\bq})-(\bj_{\bk}, \bb_{\bp}, \bu_{\bq})\} \nonumber\\
 &+& \alpha_E^b\Real\{(\bb_{\bk}, \bj_{\bp}, \bu_{\bq})+(\bj_{\bk}, \bb_{\bp}, \bu_{\bq})\}
+\beta_E^b\Real\{(\bb_{\bk}, \bb_{\bp}, \bw_{\bq})\}. \label{App:basisbb4}
\end{eqnarray}
In (\ref{identity2}) replacing $\bx, \by$ and $\bz$ by respectively $\bb$, $\bb$ and $\bu$ we have
\begin{equation} \label{App:identity3b}
 \left(\bb_{\bk}, \bj_{\bp}, \bu_{\bq}\right)  + \left(\bj_{\bk}, \bb_{\bp}, \bu_{\bq}\right)=\left(\bb_{\bk}, \bb_{\bp}, \bw_{\bq}\right)+ 2\ii(\bp\cdot\bu_{\bq})( \bb_{\bp}\cdot\bb_{\bk}).
\end{equation}
Then replacing (\ref{App:identity3b}) in (\ref{App:basisbb4}) leads to
\begin{eqnarray}
T_E(\bb_\bp\overset{\bu_\bq}{\longrightarrow}\bb_\bk) &=&\Imag\{(\bp\cdot\bu_{\bq})( \bb_{\bp}\cdot\bb_{\bk})\} +\Delta_E^b(\bk|\bp|\bq),\label{App:basisbb5}
\end{eqnarray}
with
\begin{equation}
\Delta_E^b(\bk|\bp|\bq)=\alpha_E^b\Real\{(\bb_{\bk}, \bj_{\bp}, \bu_{\bq})+(\bj_{\bk}, \bb_{\bp}, \bu_{\bq})\}
+\beta_E^b\Real\{(\bb_{\bk}, \bb_{\bp}, \bw_{\bq})\},\label{App:DeltaEb2}
\end{equation}
which is (\ref{Tbb}) with (\ref{eq:DeltaEb}).
Finally (\ref{Tbu}) results from (\ref{Tub}) exchanging $\bp$ and $\bk$.

\subsection{Derivation of kinetic helicity mode-to-mode transfer (\ref{TuuH})}
\label{sec:derivation mtmt kinetic helicity}
From (\ref{eq:SversusT}) we have
\begin{eqnarray}
S_H^u(\bk|\bp,\bq) &=& T_H(\bu_\bp\overset{\bu_\bq}{\longrightarrow}\bu_\bk)+T_H(\bu_\bq\overset{\bu_\bp}{\longrightarrow}\bu_\bk), \label{App:SuversusTHu}
\end{eqnarray}
with $S_H^u$ defined in (\ref{SH1}) as
\begin{equation}
S_H^u\left(\bk|\bp,\bq\right)=\Real\{(\bw_\bk,\bu_\bp,\bw_\bq)-(\bw_\bk,\bw_\bp,\bu_\bq)\},\label{App:SH1}\\
\end{equation}
and $T_H(\bu_\bp\overset{\bu_\bq}{\longrightarrow}\bu_\bk)$ and $T_H(\bu_\bq\overset{\bu_\bp}{\longrightarrow}\bu_\bk)$ defined from (\ref{basisuuH}) as
\begin{eqnarray}
T_H(\bu_\bp\overset{\bu_\bq}{\longrightarrow}\bu_\bk) &=& A_4\Real\{(\bw_{\bk}, \bw_{\bp}, \bu_{\bq})\}
+ B_4\Real\{(\bw_{\bk}, \bu_{\bp}, \bw_{\bq})\}
+C_4\Real\{(\bu_{\bk}, \bw_{\bp}, \bw_{\bq})\},\;\;\;\;\;\;\;\;\label{App:basisuuH}\\
T_H(\bu_\bq\overset{\bu_\bp}{\longrightarrow}\bu_\bk) &=& A_4\Real\{(\bw_{\bk}, \bw_{\bq}, \bu_{\bp})\}
+ B_4\Real\{(\bw_{\bk}, \bu_{\bq}, \bw_{\bp})\}
+C_4\Real\{(\bu_{\bk}, \bw_{\bq}, \bw_{\bp})\}.
\label{App:basisuuH2}
\end{eqnarray}
Replacing (\ref{App:SH1}), (\ref{App:basisuuH}) and (\ref{App:basisuuH2}) in (\ref{App:SuversusTHu}) leads to $B_4-A_4=1$.

From (\ref{eq:Topposite}) we have
\begin{eqnarray}
\label{App:ToppositeH}
T_{H}(\bu_\bp\overset{\bu_\bq}{\longrightarrow}\bu_\bk)&=&
-T_{H}(\bu_\bk\overset{\bu_\bq}{\longrightarrow}\bu_\bp).
\end{eqnarray}
with
\begin{eqnarray}
T_H(\bu_\bk\overset{\bu_\bq}{\longrightarrow}\bu_\bp) &=& A_4\Real\{(\bw_{\bp}, \bw_{\bk}, \bu_{\bq})\}
+ B_4\Real\{(\bw_{\bp}, \bu_{\bk}, \bw_{\bq})\}
+C_4\Real\{(\bu_{\bp}, \bw_{\bk}, \bw_{\bq})\},\;\;\;\;\;\;\;\;\;\label{App:basisuuH3}
\end{eqnarray}
Replacing (\ref{App:basisuuH}) and (\ref{App:basisuuH3}) in (\ref{App:ToppositeH}) leads to $B_4=C_4$.

Taking $A_4=\alpha_H^u-1$, $B_4=C_4=\alpha_H^u$ in (\ref{App:basisuuH})  leads to 
\begin{eqnarray}
T_E(\bu_\bp\overset{\bu_\bq}{\longrightarrow}\bu_\bk) &=& 
-\Real\{(\bw_{\bk}, \bw_{\bp}, \bu_{\bq})\} +
\Delta_H^u(\bk|\bp|\bq),\label{App:basisuuH4}
\end{eqnarray}
with
\begin{equation}
\Delta_H^u(\bk|\bp|\bq)=\alpha_H^u\Real\{(\bw_{\bk}, \bw_{\bp}, \bu_{\bq})
+ (\bw_{\bk}, \bu_{\bp}, \bw_{\bq})
+(\bu_{\bk}, \bw_{\bp}, \bw_{\bq})\},\label{App:DeltaHu}
\end{equation}
which is (\ref{TuuH}) with (\ref{eq:DeltaHu}).

\section{Derivation of fluxes}
\subsection{Derivation of equation (\ref{eq:flux6})}
\label{sec:derivation Fluxes}
From (\ref{eq:flux}) we have
\begin{equation}
\Pi^{y<}_{x>}(k_0)=\sum_{p\le k_0}\sum_{k>k_0}T_E(\by_\bp\overset{\bz_\bq}{\longrightarrow}\bx_\bk) \label{Ap:eq:flux}
\end{equation}
that can be split into the sum of two terms for $q\le k_0$ and $q>k_0$
\begin{equation}
\Pi^{y<}_{x>}(k_0)=\sum_{p\le k_0}\sum_{q\le k_0}\sum_{k>k_0}T_E(\by_\bp\overset{\bz_\bq}{\longrightarrow}\bx_\bk)
+ \sum_{p\le k_0}\sum_{q> k_0}\sum_{k>k_0}T_E(\by_\bp\overset{\bz_\bq}{\longrightarrow}\bx_\bk). \label{Ap:eq:flux2}
\end{equation}
Exchanging $p$ and $k$ in the second term on the right hand side of (\ref{Ap:eq:flux2}) leads to
\begin{equation}
\Pi^{y<}_{x>}(k_0)=\sum_{p\le k_0}\sum_{q\le k_0}\sum_{k>k_0}T_E(\by_\bp\overset{\bz_\bq}{\longrightarrow}\bx_\bk)
+ \sum_{k\le k_0}\sum_{q> k_0}\sum_{p>k_0}T_E(\by_\bk\overset{\bz_\bq}{\longrightarrow}\bx_\bp)\label{Ap:eq:flux3}
\end{equation}
which, using (\ref{eq:Topposite}), is equivalent to
\begin{equation}
\Pi^{y<}_{x>}(k_0)=\sum_{p\le k_0}\sum_{q\le k_0}\sum_{k>k_0}T_E(\by_\bp\overset{\bz_\bq}{\longrightarrow}\bx_\bk)
- \sum_{p>k_0}\sum_{q> k_0}\sum_{k\le k_0}T_E(\bx_\bp\overset{\bz_\bq}{\longrightarrow}\by_\bk).\label{Ap:eq:flux4}
\end{equation}

As (\ref{Ap:eq:flux4}) is invariant under the exchange of $\bp$ and $\bq$ it can be written in the form
\begin{eqnarray}
\Pi^{y<}_{x>}(k_0)&=&\frac{1}{2}\sum_{p\le k_0}\sum_{q\le k_0}\sum_{k>k_0}T_E(\by_\bp\overset{\bz_\bq}{\longrightarrow}\bx_\bk)+T_E(\by_\bq\overset{\bz_\bp}{\longrightarrow}\bx_\bk)\nonumber \\
&-& \frac{1}{2}\sum_{p>k_0}\sum_{q> k_0}\sum_{k\le k_0}T_E(\bx_\bp\overset{\bz_\bq}{\longrightarrow}\by_\bk)+T_E(\bx_\bq\overset{\bz_\bp}{\longrightarrow}\by_\bk)\label{Ap:eq:flux5}
\end{eqnarray}
leading to
\begin{eqnarray}
\Pi^{y<}_{x>}(k_0)&=&\frac{1}{2}\sum_{p\le k_0}\sum_{q\le k_0}\sum_{k>k_0}S^{xy}_E(\bk|\bp,\bq) 
- \frac{1}{2}\sum_{p>k_0}\sum_{q> k_0}\sum_{k\le k_0}S^{yx}_E(\bk|\bp,\bq) \label{Ap:eq:flux6}
\end{eqnarray}
which is (\ref{eq:flux6}).

\subsection{Derivation of (\ref{eq:sumfluxcste}) from (\ref{eq:EbfromT})}
\label{App:Flux Identity}
Rewriting (\ref{eq:EbfromT}) we have
\begin{equation}
( \partial_t  + 2\eta k^2) E^b_\bk  = \sum_\bp
T_E(\bu_\bp\overset{\bb_\bq}{\longrightarrow}\bb_\bk)+T_E(\bb_\bp\overset{\bu_\bq}{\longrightarrow}\bb_\bk),
\label{App:eq:EbfromT}
\end{equation}
with $\bf k+p+q=0$.
Taking the sum of (\ref{App:eq:EbfromT}) over all $\bk$ is equivalent to
\begin{equation}
\sum_\bk( \partial_t  + 2\eta k^2) E^b_\bk  = \sum_\bk\sum_\bp
T_E(\bu_\bp\overset{\bb_\bq}{\longrightarrow}\bb_\bk)+T_E(\bb_\bp\overset{\bu_\bq}{\longrightarrow}\bb_\bk).
\label{App:eq:EbfromT2}
\end{equation}
Splitting the sums as $\sum  \limits_\bk=\sum \limits_{\underset{k\le k_0}{\bk}}+\sum \limits_{\underset{k > k_0}{\bk}}$
and $\sum  \limits_\bp=\sum \limits_{\underset{p\le k_0}{\bp}}+\sum \limits_{\underset{p > k_0}{\bp}}$,
we have
\begin{equation}
\sum_\bk\sum_\bp
T_E(\bu_\bp\overset{\bb_\bq}{\longrightarrow}\bb_\bk)= \Pi^{u<}_{b<}(k_0)+\Pi^{u<}_{b>}(k_0)+\Pi^{u>}_{b<}(k_0)+\Pi^{u>}_{b>}(k_0)
\label{App:eq:EbfromT3}
\end{equation}
and
\begin{equation}
\sum_\bk\sum_\bp
T_E(\bb_\bp\overset{\bu_\bq}{\longrightarrow}\bb_\bk)= \Pi^{b<}_{b<}(k_0)+\Pi^{b<}_{b>}(k_0)+\Pi^{b>}_{b<}(k_0)+\Pi^{b>}_{b>}(k_0).
\label{App:eq:EbfromT4}
\end{equation}
From (\ref{eq:flux})  we have for $\by=\bx$
\begin{equation}
\Pi^{x<}_{x>}(k_0)=\sum_{p\le k_0}\sum_{k> k_0}T_E(\bx_\bp\overset{\bz_\bq}{\longrightarrow}\bx_\bk) .\label{Ap:eq:flux7}
\end{equation}
Exchanging $p$ and $k$ leads to
\begin{equation}
\Pi^{x<}_{x>}(k_0)=\sum_{k\le k_0}\sum_{p> k_0}T_E(\bx_\bk\overset{\bz_\bq}{\longrightarrow}\bx_\bp) .\label{Ap:eq:flux8}
\end{equation}
which, using (\ref{eq:Topposite}), gives
\begin{eqnarray}
\Pi^{x<}_{x>}(k_0)&=&-\sum_{k\le k_0}\sum_{p> k_0}T_E(\bx_\bp\overset{\bz_\bq}{\longrightarrow}\bx_\bk) \label{Ap:eq:flux9}\\
						&=&-\Pi^{x>}_{x<}(k_0). \label{Ap:eq:flux10}
\end{eqnarray}
Therefore in (\ref{App:eq:EbfromT4}) we have $\Pi^{b<}_{b>}(k_0)+\Pi^{b>}_{b<}(k_0)=0$. 

Similarly we can demonstrate taking $p\le k_0$ and $k\le k_0$
that $\Pi^{x<}_{x<}(k_0)=0$, taking $p> k_0$ and $k> k_0$
that $\Pi^{x>}_{x>}(k_0)=0$. Therefore in (\ref{App:eq:EbfromT4}) we have $\Pi^{b<}_{b<}(k_0)=\Pi^{b>}_{b>}(k_0)=0$, implying that $\sum \limits_\bk\sum \limits_\bp T_E(\bb_\bp\overset{\bu_\bq}{\longrightarrow}\bb_\bk)=0$.

Then (\ref{App:eq:EbfromT2}) can be written as
\begin{equation}
\sum_\bk( \partial_t  + 2\eta k^2) E^b_\bk  = \Pi^{u<}_{b<}(k_0)+\Pi^{u<}_{b>}(k_0)+\Pi^{u>}_{b<}(k_0)+\Pi^{u>}_{b>}(k_0).
\label{App:eq:EbfromT5}
\end{equation}
As the left hand side of (\ref{App:eq:EbfromT5}) does not depend on $k_0$ its derivative with respect to $k_0$ is zero, leading to (\ref{eq:sumfluxcste}).

\end{document}